\documentclass[apj]{emulateapj}

\slugcomment{Accepted for publication in ApJ}

\shorttitle{The Enrichment of the ICM}

\begin{document}

\title{The Enrichment of the Intracluster Medium}

\author{Suresh Sivanandam\altaffilmark{1}, Ann I. Zabludoff\altaffilmark{1}, Dennis Zaritsky\altaffilmark{1}, Anthony H. Gonzalez\altaffilmark{2}, and Dan D. Kelson\altaffilmark{3}}

\altaffiltext{1}{Steward Observatory, University of Arizona, 933 North Cherry Ave, Tucson, AZ 85721; suresh@as.arizona.edu, azabludoff@as.arizona.edu, dzaritsky@as.arizona.edu.}

\altaffiltext{2}{Department of Astronomy, University of Florida, P.O. Box 112055, 211 Bryant Space Science Center, Gainesville, FL 32611; anthony@astro.ufl.edu.}

\altaffiltext{3}{The Observatories, Carnegie Institution of Washington, 813 Santa Barbara St, Pasadena, CA 91101; kelson@ociw.edu.}

\begin{abstract}
To determine the relative contributions of galactic and intracluster stars to the enrichment of the intracluster medium (ICM), we present X-ray surface brightness, temperature, and Fe abundance profiles for a set of twelve galaxy clusters\footnotemark[4] for which we have extensive optical photometry. Assuming a standard IMF and simple chemical evolution model scaled to match the present-day cluster early-type SN Ia rate, the stars in the brightest cluster galaxy (BCG) plus the intracluster stars (ICS)  generate $31^{+11}_{-9}$\%, on average, of the observed ICM Fe within r$_{500}$ ($\sim 0.6$ times r$_{200}$, the virial radius). An alternate, two-component SN Ia model (including both prompt and delayed detonations) produces a similar BCG+ICS contribution of $22^{+9}_{-9}$\%. Because the ICS typically contribute 80\% of the BCG+ICS Fe, we conclude that the ICS are significant, yet often neglected, contributors to the ICM Fe within r$_{500}$. However, the BCG+ICS fall short of producing all the Fe, so metal loss from stars in other cluster galaxies must also contribute. By combining the enrichment from intracluster and galactic stars, we can account for all the observed Fe.  These models require a galactic metal loss fraction ($0.84^{+0.11}_{-0.14}$) that, while large, is consistent with the metal mass not retained by galactic stars. The SN Ia rates, especially as a function of galaxy environment and redshift, remain a significant source of uncertainty in further constraining the metal loss fraction.  For example, increasing the SN Ia rate by a factor of 1.8 ---  to just within the 2$\sigma$ uncertainty for present-day cluster early-type galaxies ---  allows the combined BCG + ICS + cluster galaxy model to generate all the ICM Fe with a much lower galactic metal loss fraction ($\sim 0.35$).

\footnotetext[4]{Based on observations obtained with XMM-Newton, an ESA science mission with instruments and contributions directly funded by ESA Member States and NASA.}

\end{abstract}

\keywords{galaxies: clusters: general -- X-rays: galaxies: clusters -- galaxies: intergalactic medium -- stars: supernovae: general}

\section{Introduction}
The high measured Fe abundance of the intracluster medium (ICM) of galaxy clusters, $\sim 0.3$ Z$_{Fe,\odot}$ \citep{edge}, has been challenging to explain. The source of these metals is still unknown. Various enrichment mechanisms have been proposed, including models in which cluster galaxies lose enriched gas via galactic winds \citep{young}, ram-pressure stripping \citep{gunn}, or gravitational interactions \citep{moore}, but these have fallen short of producing the observed metals unless non-standard assumptions are made; for example, top-heavy initial mass functions (IMF) \citep{mushotzky,gibson97}, severe mass loss from galaxies \citep{renzini93,renzini97}, or pre-enrichment from Population III hypernovae \citep{lowenstein01}. 

Such exotic explanations are intriguing, but we must first rule out more mundane explanations. 
One source of ICM metals that has only recently been considered is
the diffuse, stellar component in clusters, {\it i.e.}, the stars that lie outside the member galaxies \citep{domainko04,zaritsky04,linmohr}. These stars must be a source of metals for the ICM because they pollute {\it in-situ}; the key
question is whether they are a significant, or perhaps even the dominant, source. 

Recent measurements of the intracluster stars (ICS) in a large sample of clusters 
demonstrate that this component contains roughly 30\% of the stellar mass out to r$_{500}$\footnote{The radius within which the mass overdensity is 500 times the universal value and which
corresponds to $\sim$60\% of the virial radius \citep{cole96}.} 
and an even larger
percentage at the smaller radii typically probed by X-ray measurements \citep{gonzalez07}. 
In previous work, \citet{gonzalez05} detect
these stars in every cluster in their sample out to 
a radius of at least 300 kpc ($h=0.7$).  By stacking SDSS cluster images,  \citet{zibetti} detect the averaged ICS
to 700 kpc, demonstrating the large extent of this component.
Without any further calculation, one 
concludes that
these stars 
could contribute 
$\sim$ 30\% of the ICM's Fe.

Our first test
of this scenario was promising, suggesting that intracluster stars could play a significant role in ICM enrichment \citep{zaritsky04}. This result was preliminary because we lacked spatially-resolved metallicity profiles to match apertures between the optical and X-ray measurements, and we had the minimum necessary data to make the optical/X-ray comparisons for just three clusters, of which only two had intracluster light measurements based on our high sensitivity methods. Because intracluster stars are typically more concentrated than the cluster galaxies \citep{gonzalez07}, and clusters often have radial abundance gradients \citep{degrandi}, 
our test 
must be re-done using matched apertures and homogeneous data.

We have measured uniformly
the properties of the intracluster stellar components for a sample of 28 clusters (24 of which are from \citet{gonzalez05}, while the remainder are from \citet{kelson}). In this paper, we use XMM-Newton archival and guest observer data to 
measure the radial distribution of the X-ray emitting plasma and the Fe within that plasma for those
12 clusters with available X-ray data. We then determine the gas and Fe masses enclosed within the
same radii used for the stellar mass measurements.

The outline of this paper is as follows. In $\S$2 we review our sample. In $\S$3 we discuss the data and our analysis. In $\S$4 we present the X-ray spectral, morphological, and gas mass measurements. In $\S$5 we discuss the comparison of the observed Fe abundances with those predicted by a simple chemical evolution model assuming metal enrichment from the intracluster and galactic stars. Finally, in $\S$6 we summarize our conclusions. Throughout this paper, except where comparing to previous work and noted, we adopt the concordance cosmological model ($\Omega_\Lambda = 0.70,$ $\Omega_m =
0.30,$ and $H_0 = 70$ km s$^{-1}$ Mpc$^{-1}$), which we call LCDM70 and define r$_X$ to be
the radius of a sphere within which the mass density is $X$ times the universal mean. All reported errors in this work are quoted at the 1$\sigma$ level.

\section{Sample}
The clusters in this study are a subset of those for which we have intracluster light measurements \citep{gonzalez05,kelson}. First, we searched the XMM-Newton Science Archive (XSA) for European Photon Imaging Camera (EPIC) observations of any of these clusters. Observations of eight clusters are available, although one, Abell 3376, is not included here due to its highly disturbed X-ray morphology. Second, we obtained XMM-Newton Guest Observer (GO) observations of five additional clusters. The selection criteria for the sample of twelve clusters presented here (see Table \ref{sourcelist}) are a 
combination of
 those in our original intracluster stellar light work, 
those by which others selected clusters for X-ray observations, and 
those we imposed in
selecting targets for GO observations.

Our original intracluster light catalog comes from two sources and so itself has varied selection criteria: \citet{gonzalez05} chose their clusters to have a single dominant brightest cluster galaxy (BCG) with a major axis position angle within 45$^\circ$ of east-west (to accommodate their drift-scan exposures) and no visual evidence of an ongoing merger; \citet{kelson} chose their clusters to have a BCG that is kinematically stationary relative to the cluster, to appear relaxed in X-rays (when such data were available), and to be at sufficiently large redshift to not overfill the spectrograph
slit length and close enough so that the kinematics of the low surface brightness ICS could be measured.

Although we are unable to reconstruct the selection criteria for the X-ray archival sources, there is a natural tendency in the archival data to sample more X-ray luminous systems and Bautz-Morgan (BM) \citep{bautz} Type I clusters. For our GO observations, our aim was to observe more of the \citet{gonzalez05} clusters, for which clear distinctions between the BCG and ICS components has been measured, allowing us to make better optical/X-ray comparisons.

The selection criteria described above do not lead to a representative sample of clusters. Instead, all clusters have a central BCG (BM type I or I-II with Abell 3705 being the sole BM type III), Abell cluster richness class values of 1 or 2 \citep{abell58}, and show no clear optical evidence of an ongoing major merger. The clusters are within the redshift range $0.02 < z < 0.13$ and have line-of-sight velocity dispersions, $\sigma$, spanning 500 to 1000 km s$^{-1}$, corresponding to the masses of rich groups and clusters ($\lesssim 10^{14} - 10^{15}$ M$_\odot$). We determine the size of these clusters by 
associating r$_{200}$, the radius within which the mass overdensity is
200 times the universal value, with the virial radius \citep{cole96}.
In practice, we measure r$_{200}$ only for the \cite{gonzalez05} clusters, for which
we have measured galaxy overdensities \citep[see][for details]{gonzalez07}.  We then
use the theoretical relation between r$_{200}$ and $\sigma$ \citep{finn05}, renormalized to match the \citet{gonzalez07} r$_{200}$ values, for the \cite{kelson} clusters. We use the r$_{500}-\sigma$ relation\footnote{All clusters in this paper lie in the $\sigma$ range ($\sim 500$ to 1100 km s$^{-1}$)  where the r$_{500}-\sigma$ relation is calibrated. We advise against 
extrapolating this relation to lower velocity dispersions \citep[cf.][]{balogh08}} derived in \cite{gonzalez07} to compute the r$_{500}$ values for all our clusters. We present the physical parameters of the clusters in Table \ref{sourcelist}. The quoted cluster temperatures are the maximum annularly-averaged temperatures observed in our spectral fits (see $\S$4.6 for details). This choice mitigates the ambiguity of the global temperature for cool core clusters and better reflects the depth of the cluster potential.

\section{Data and Analysis}
\subsection{X-ray Data: Preprocessing}

For our sample of 12 clusters, we have XMM-Newton EPIC X-ray ($0.1-12$ keV) observations, optical photometry ($i$-band from \citet{gonzalez05} or R-band from \citet{kelson}), and optical spectroscopy of member galaxies \citep{zaritsky05}. In this section, we focus on the analysis of the X-ray data.

We present in Table \ref{dataproperty} the observation ID of each X-ray data file, the date of observation, the source of the data, the filter used, EPIC-MOS camera exposure times, and the radius, r$_{\mathrm{outer}}$, out to which we perform spatially resolved spectroscopy. The exposure times quoted are the average exposure times of the MOS1 and MOS2 imaging spectrographs. We provide both the original unfiltered exposure time t$_{\mathrm{MOS}}$ and the final flare-filtered exposure time t$_{\mathrm{MOS,ff}}$ for each exposure. All MOS exposures were taken in Full Frame mode.

We only analyze the MOS data because XMM-ESAS, a quiescent background modeling software that effectively models the time-variable instrument and particle background spectra, is only available for MOS and because the XMM reflection grating spectrograph (RGS) does not offer the necessary field-of-view or sensitivity. We use XMM Science Analysis Software (SAS) 6.5 and FTOOLS 5.3.1 to reduce the MOS data and create any XMM calibration or ancillary data.

The MOS data are prepared for further analysis as follows. We run the \emph{EMCHAIN} task on the original data products to generate the event lists for each observation. We use recent calibration files (January 2006), which take into account both the spatial and temporal variability of the MOS response \citep{stuhlinger06}. We choose all events, without any restriction on the energy range, with (PATTERN $\leq$ 12) to filter out cosmic rays and other non-X-ray events. We set ((FLAG \& 0x766a0f63) == 0) to remove any bad pixels and retain the events that fall outside the field-of-view, namely in the unilluminated ``corner" pixels used later for the background analysis.

\subsection{Background Subtraction}
Various sources contribute background to the observations: 1) soft proton flares, which render the data coincident with the flare useless, 2) a quiescent low-energy particle background, which exists even when no flares are present and is independent of the observed field, and 3) a ``cosmological" background, which consists of resolved and unresolved sources in the field of view (for a complete review see \cite{snowden07}).

XMM-Newton observations are plagued by flaring problems from the satellite's passage through the Earth's radiation belt and proton clouds. These flares are typically caused by soft protons that are funneled into the telescope through the focusing optics and impinge on the detector. We remove these transient events before further analysis by excising all events for which the 2.5 to 8.5 keV count rate exceeds a specified threshold within a time interval of 1 second. We determine the appropriate threshold value by generating a histogram of the count rate, fitting a Gaussian to the distribution to obtain the mean and dispersion $\sigma_{ctr}$, and identifying the count rate that is 2.5$\sigma_{ctr}$ above the mean. We remove events in time intervals where the rate exceeds this threshold and iterate this procedure twice. The final integration times after flare-filtering are given in Table \ref{dataproperty}. This filtration is fairly rudimentary because it is insensitive to low intensity proton flares. We account for this transient component in our spectral fitting (see $\S  $3.4 for details). In general, we find $>50\%$ of the integration time of any given X-ray observation is useable for further analysis.

The second background source consists of X-rays generated within the instrument by the constant flux of low energy protons impinging on the detector and the surrounding housing. This background, both in the form of spectral lines due to fluorescence (Al and Si K$\alpha$) and continuum, changes temporally and spectrally, and is present even when no flaring is observed. All observations must be corrected for it. We use the techniques outlined by \citet{kuntz04} and implemented in the XMM-ESAS package for the quiescent particle background subtraction.

The procedure outlined by \citet{kuntz04} is a step-up in sophistication from typical quiescent background subtraction techniques, which simply scale filter-wheel closed (FWC) data to match the corner pixel (i.e., dark) count rates. The quiescent particle background is known to vary temporally in spectral shape and in strength over long time scales (i.e., over several orbits). Corner pixels, which are not exposed to photons because they lie outside of the field-of-view, are exposed only to this background over the same time interval as the observation. \citet{kuntz04} characterize the observed corner pixel spectral using two spectral parameters: the high energy (2.5 to 12.0 keV) power law slope and the hardness ratio, which is the $2.5-5.0$ keV to $0.4-0.8$ keV band flux ratio.

There are too few corner pixel counts to generate a background spectrum with sufficient S/N.
Their method uses a database of FWC observations and extracts corner pixel data with similar slopes and hardness ratios for each MOS chip to augment and improve the S/N of the corner pixel spectrum of the observation. The background spectrum within the region of interest is created by multiplying the corner spectrum derived for each chip by the ratio of FWC object region spectrum and the FWC corner pixel spectrum. Lastly, for a specified source region they combine the individual background components derived for each of the 7 chips using appropriate weights based on each chip's coverage of the source region and create a single background spectrum.

We discuss our methods for addressing the ``cosmological'' background in the next section.

\subsection{X-ray Data: Spectroscopy}
To measure abundance and temperature profiles, we generate spatially resolved spectra for circular annuli centered on the clusters. We first identify the cluster center, set the annuli, locate point sources within each annulus, and remove point sources. We find the cluster center on a surface brightness map. Then, we radially bin the data in multiple annuli of variable widths ranging from 30 to 300 arcsec to obtain the highest radial resolution while maximizing the signal-to-noise in each bin. We choose 30 arcsec as a lower limit to minimize PSF blending effects between adjacent bins. The binning choice varies for each cluster and is a function of the signal-to-noise of a particular observation. To remove point sources, we generate an image of each exposure and use the CIAO (Chandra Interactive Analysis of Observations) 3.3 wavelet source detection utility, \emph{WAVDETECT}. For each source, we obtain from \emph{WAVDETECT} the elliptical aperture that contains 99.73\% (within 3$\sigma$ of a Gaussian distribution) of the source light. Elliptical apertures are necessary due to the off-axis changes in the XMM-Newton PSF. We visually inspect each source list and associated elliptical apertures to remove any spurious detections and ensure that the source extraction regions are large enough. We generate a region file containing the acceptable (point-source free) pixels in each bin.

Next, we extract spectra and correct for the particle background spectrum. Using scripts available in XMM-ESAS that implement in the \cite{kuntz04} algorithm for determining the particle background, we generate particle-background spectra that we subtract from our annularly binned cluster spectra. We compare the background and observed spectrum within an annulus to ascertain the quality of flare-filtering. At high energies ($>$10 keV) where the telescope sensitivity is effectively zero, we do not in any case find that the background spectrum significantly deviates from the observed spectrum, indicating that the flares were effectively removed. If there remains low-level flaring, which cannot be detected by the above technique, we remove it by explicitly modeling it in our spectral fits as discussed below.

Using \emph{GRPPHA}, we group all of the response and background files together per spectrum, and we bin both the observed and background spectra to have 50 counts per energy bin, which ensures Gaussian statistics. Ideally, we would want to construct energy bins to maximize signal-to-noise, but \emph{GRPPHA} does not have this feature. Count-binning is adequate for bins that are source count dominated, but inadequate for annuli that are particle background dominated. If the background counts are $>$25\% of the signal counts, the high energy tail of the background-subtracted spectrum has very few counts, which results in a noisy spectrum.

We use \emph{XSPEC 11} to fit physical models to the data and obtain measurements of temperature and chemical abundance. We fit to the $0.5-10$ keV region of the spectrum, because it covers 
the Fe L-shell ($\sim$ 1 keV) and $K_{\alpha}$ (6.7 keV) transitions, which are useful in 
determining the Fe abundance. XMM-Newton is known to have calibration issues at lower energies \citep{kirsch05} and there are too few counts at energies greater than 10 keV. 
This range is required for proper determination of the continuum level and for characterizing the instrument background \citep{snowden07}. 
We use a complete model that accounts for the source, sky background, and instrumental emission, and fit all of the annuli simultaneously. Our model of the sky background includes emission from the local hot bubble, a residual soft proton flare background, and unresolved X-ray sources. We assume the cosmic background is constant over the field-of-view. Although this assumption is not entirely correct --- the galactic X-ray background is patchy over scales of a few arcmin --- it is reasonable because the XMM-Newton field-of-view is 30 arcmin, which effectively averages out small-scale variations. To constrain the low energy end of the background, we use additional ROSAT All Sky 
Survey (RASS) data, which covers the $0.1-2.4$ keV spectral range. We use the X-ray background tool at HEASARC that culls the RASS database to obtain a ROSAT position sensitive proportional counter (PSPC) spectrum of the annular region 1 to 2 degrees from the center of the cluster and the associated PSPC response file. Most of the emission in the   $0.1-2.4$ keV range is ``background" because the cooler galactic plasma (0.1 keV) emits more efficiently at these energies than the cluster plasma ($\sim$ 2 to 10 keV), therefore the PSPC spectrum provides a critical constraint to the soft cosmic background.

We now fit the multicomponent model and follow to a large extent the prescription outlined in \cite{snowden07}. The actual fitting procedure begins with the application of the IDL \emph{findspeclim} routine to find the upper and lower spectral channels that correspond to the upper and lower energy limits ($0.5 - 10$ keV). We then run \emph{genfitfile.pl}, with known column density ($N_H$) derived from the \citet{dickey90} HI survey (using the HEASARC nH tool) and the cluster redshift from the NASA Extragalactic Database (NED) or our own data \citep{gonzalez07}, to generate an \emph{XSPEC} initial fit file. We simultaneously fit both the MOS1 and MOS2 data as well as the PSPC data with \emph{XSPEC} using the following \emph{XSPEC} model expression: {\it ``model bknpower/b + gaussian + gaussian + constant( apec + ( apec + apec + powerlaw )wabs + ( apec )wabs"}. {\it bknpower/b} is a broken power law, which is independent of the telescope's collecting area, that models any residual soft proton flares and takes the following functional form:
\begin{eqnarray}
f(E)   & = &  \kappa E^{-\alpha} \:\:\:\:\:\:\:\:\:\:\:\:\: E \leq E_0  \nonumber \\
	& = & \kappa E_0^{\beta-\alpha} E^{-\beta} \:\:\: E \geq E_0 
\end{eqnarray}
where $f(E)$ represents the flux of the model as a function of energy $E$, $\kappa$ is the normalization constant, $\alpha$ and $\beta$ are power law exponents, and $E_0$ is the break energy. We fix the break energy at 3.3 keV and allow the exponents on either side of the break to vary freely along with the model's flux normalization. We assume that the soft proton flux is constant over the instrument, and we use the data from all annuli scaled appropriately by the annular area (in units of arcmin$^2$) to constrain these parameters. The two Gaussians, signified by {\it gaussian} in the XSPEC model expression above, represented in the analytical expression below model the instrumental X-ray florescence lines: Al K$\alpha$ (1.49 keV), which is especially apparent in annuli with few cluster counts, and Si K$\alpha$ (1.74 keV). The model flux is then
\begin{eqnarray}
f(E) & = & \frac{\kappa}{\sigma\sqrt{2\pi}}\exp\left[\frac{1}{2}\left(\frac{E-E_0}{\sigma}\right)^2\right],
\end{eqnarray}
where $\kappa$ is the normalization constant, $E_0$ is the line center, and $\sigma$ is the line width.
\citet{kuntz04} have shown that the strength of these lines vary with each chip in the MOS instrument, so we fit for the flux normalization of these lines for each annulus while keeping the line energies and widths fixed. We initially also allowed the widths to vary but the best fit width values were $<$ 0.01 eV, which is less than the MOS spectral resolution. 

The remainder of the parameters model the cosmic background and the source. {\it apec} \citep{smith01} is a collisional plasma model with the following free-parameters: temperature ($T$), chemical abundance ($Z$), redshift ($z$), and flux normalization ($\kappa$). {\it wabs} \citep{morrison} implements a neutral hydrogen column photo-electric absorption model and is parameterized by $N_H.$ The {\it constant} accounts for the solid angle each annulus subtends on the sky (in arcmin$^2$). For each annulus, we fix the value of the solid angle by appropriately scaling the value computed by SAS \emph{BACKSCALE}, a utility that counts the number of square pixels within a given detector area. This is necessary to equally weight the fitting of background parameters over all annuli. The background is modeled with 4 components: the local unabsorbed hot bubble ({\it apec}, $T \sim 0.1$ keV, $\textrm{Z} = \textrm{Z}_\odot$), the cool absorbed galactic halo ({\it apec}, $T = 0.1$ keV, $\textrm{Z} = \textrm{Z}_\odot$, usually a negligible component), the warm absorbed galactic component ({\it apec}, $T \sim 0.3$ keV, $\textrm{Z} = \textrm{Z}_\odot$), and an unresolved extragalactic background. The extragalactic background is modeled as a power law in energy with an index of 1.41 and a flux normalization of $8.88\times10^{-7}$ counts sec$^{-1}$ keV$^{-1}$ arcmin$^{-2}.$ We fit the neutral hydrogen column of the absorbed models using the HI column density observed at the location of the cluster \citep{dickey90} as an initial guess. Lastly, the cluster emission itself is modeled with an {\it apec} model. For each annulus, all parameters for both MOS1 and MOS2 spectra are tied together. We fix the redshift of the cluster to that in NED or to our own measured value \citep{gonzalez07}. We adopt \citet{anders89} solar photospheric abundances. We use $\chi^2$ minimization to obtain best fit values. The fits provide a measure of the cluster temperature, Fe abundance, and flux normalization for each annulus. The maximum radius of our spectral fits, r$_{spec}$, given in Table \ref{gasmass} is determined to be the radius within which the fractional error of the derived cluster abundance does not exceed 50\%.

\subsection{X-ray Data: Imaging}

\subsubsection{Generation of Smoothed Maps}

Our principal objective for the imaging is to examine the cluster morphologies, generate surface brightness profiles, and perform surface brightness fits. We take the  MOS1 and MOS2 flare-filtered event lists of each target and generate a soft ($0.3 - 1.25$ keV) and a hard ($2 - 8$ keV) image. We choose these energy ranges to avoid contamination from the Al and Si K$\alpha$ fluorescence lines. We generate exposure maps with \emph{EEXPMAP} for each image, which are used to flatten the data and remove vignetting. We create bad pixel masks using \emph{EMASK}. We identify bad columns by searching for steep gradients in the exposure map and remove them using the gradient parameter in \emph{EMASK}. Using the source region files produced for spectral fitting, we mask out point sources. For each image, we use \emph{XMM-BACK-IMAGE} to generate a corresponding particle background image. We mosaic the background-subtracted images to form a single $0.3-8$ keV image. If the data at a given sky position are available in multiple exposures, we add the fluxes. If data are available at a given sky position from only one pixel, we scale the flux up to the total exposure time. If data are not available at a given sky position, i.e., if the corresponding pixel is masked in both MOS exposures, we set the image value to zero.

To produce higher signal-to-noise images, we adaptively smooth the mosaicked image using \emph{ASMOOTH}. Our goal is signal-to-noise $\sim$ 20 over the cluster, and so we choose a maximum kernel width of 10 pixels (25 arcsec), no minimum size, and ignore image values of zero. We present the smoothed images in Figure \ref{contourfig1}. Contours are spaced logarithmically and the lowest contour is set at a fixed value of $2.0\times10^{-6}$ counts s$^{-1}$ arcsec$^{-2}$ chosen to be unaffected by the noise present in the smoothed image.

\subsubsection{Surface Brightness Fitting}
\label{sbfitting}
To measure and model the Fe mass, we require a measurement of the gas density profile. We derive this density profile from the 2D surface brightness profile fitting of MOS images and the spectral fits. The surface brightness profile constrains the shape of the gas density distribution, whereas the spectral fits provide the normalization for that distribution. We use \emph{SHERPA}, a versatile fitting package that is part of the CIAO 3.3 distribution, to fit the 2D model. Starting with the particle-background-subtracted, exposure-corrected images, we prepare these images for fitting by first adding a small offset of $1.6\times10^{-6}$ counts s$^ {-1}$ arcsec$^{-2}$ for the soft band and $3.2\times10^{-6}$ counts s$^ {-1}$ arcsec$^{-2}$ for the hard band, which excludes negative pixels. We then multiply the images with an unvignetted exposure map and add the particle background to the image. For this fit, we use the unvignetted exposure maps with point source regions zeroed out, and convolve the model with the 2D point spread function (PSF) generated by \emph{CALVIEW}. Because we are fitting an extended source, off-axis PSF effects are not as important, so we use the on-axis PSF over the entire field. For simplicity and consistency with previous studies, we choose to fit the canonical beta model \citep{cavaliere78}, which is a straightforward relation between surface brightness $I_X(R)$ and volume density $\rho_g(r)$:
 \begin{equation}
I_X(R) = I_X(0) \left[1+\left(\frac{R}{r_X}\right)^2\right]^{-3\beta+1/2}
\end{equation}
and
\begin{equation}
\rho_g(r) = \rho_g(0) \left[1+\left(\frac{r}{r_X}\right)^2\right]^{-3\beta/2},
\label{gasbeta}
\end{equation}
where the free parameters $I_X(0)$ and $\rho_g(0)$ are model normalizations, $r_X$ is the core radius, and $\beta$ defines the rate at which these profiles decline with projected 
radius $R$ and unprojected radius $r$, respectively.

We model the background, which includes both the vignetting-corrected sky and unvignetted particle background, with a constant term. This approach produces consistent results with fits performed using blank-sky subtracted images in two representative fields. For the blank-sky subtraction, we use the \cite{carter07} blank-sky files that were observed though the same filter as our observations. These files include both the particle and cosmic backgrounds. We cast the blank-sky event list files into our sky 
coordinates and scale them to match the exposure time of our exposures before carrying out the background subtraction.  

For our fits, we simultaneously fit both the soft and hard band data, using the respective unvignetted exposure maps. We use the Cash statistic, which is appropriate at low count levels, to evaluate goodness-of-fit and the simplex algorithm to find the best fit parameters. We look at residuals to determine if there are clusters with bright cores. In clusters with unusually bright centers (see Appendix for more details), we mask the central $\sim 30$\arcsec \ and refit the data. The best fit for Abell 496 converges to one with zero background flux, which is unphysical. We attribute this to large residual non-axis symmetric features found in the inner 4\arcmin \ of the cluster (see Figure \ref{residualfig1}). 
Such anomalies are rare and are discussed in the Appendix.

Initially, using  a simple spherically symmetric beta model, we fit for the center, core radius, $\beta$, and constant background level. The best fit had large residuals in most cluster centers, (see Figure \ref{2dresiduals} for one representative case), which suggested that we consider more complex models. Therefore, we included variable ellipticity and position angle. This change improved the fits, and the resulting residuals were no longer globally asymmetric; however, the statistical improvement was marginal. Because cluster X-ray emission can have two temperature components (a lower temperature, higher density component near the core, and a higher temperature, lower density component at larger radii; \cite{jia04}), we then explored fitting two elliptical beta models. The two component model performed only marginally better than the single component model and had poorly constrained outer component parameters as well as $\beta$ values that were unphysically large $(>2)$. We therefore adopt our single component, spherical beta model fits to maintain consistency with our spectral analyses, which use circular and not elliptical annuli, and to simplify our mass calculation.

To account for the fact that we are observing surface densities but want to work with volume densities, we must deproject the observations to normalize the density model.  To do this deprojection, we model the cluster as spherically symmetric with a gas density profile that obeys Equation \ref{gasbeta}.  We calculate the gas density using the Abell transform and compute the central gas density that reproduces the derived cluster {\it apec} normalization constant $\kappa$ (given below) for each annulus, assuming $n_H = 0.82 n_e$ \citep{pizzolato03}: 
\begin{equation}
\kappa = \frac{10^{-14}}{4\pi (D_A(1+z))^2}\int n_e n_H dV,
\end{equation}
where $D_A$ is the angular diameter distance (in cm), $z$ is the redshift, $n_e$ and $n_H$ are the electron density and the Hydrogen density in cm$^{-3}$, respectively. We average the central gas densities derived from fitting to individual annuli for use as the central gas density $\rho_H(0)$. We compute masses by integrating Equation \ref{gasbeta}.

\subsection{Monte Carlo Error Modeling}
We determine all central density and cluster mass errors through Monte Carlo simulations. We adopt the  spectral and radial profile best fit values and the 1$\sigma$ errors derived by \emph{XSPEC} and \emph{SHERPA}. The error distribution is modeled as Gaussian. \emph{SHERPA}'s projection function carries out its own proper Monte Carlo simulation for a given fit and returns asymmetric 1$\sigma$ errors. We model the asymmetric errors with two Gaussians. We carry out 2000 trials per cluster where for each trial we generate new values of spectral normalization, abundance, $\beta$, and $r_X$, and compute the masses. The resulting distribution of derived masses is slightly asymmetric, but centrally peaked. Increasing the number of trials does not change the resulting distribution. Using the resulting distribution of mass estimates, we evaluate our 1$\sigma$ confidence limits.

\subsection{X-ray Luminosities}
We measure cluster X-ray luminosities 
by combining the beta profile and spectral fits. 
Therefore, we can only do this for
clusters
where we were able to fit a beta profile, which excludes Abell 2877 and 3705. We first compute the luminosity enclosed within the radius of our spectral fit by using the normalization parameters of the {\it apec} model for each annulus. To determine the luminosity lying outside of the spectral fit radius, we extrapolate the luminosity out to r$_{200}$ using the beta model fit. To accurately quantify the errors in this luminosity, we carry out a Monte Carlo simulation using \emph{XSPEC 12}. We first load the best fit spectral model into \emph{XSPEC} and compute new parameters drawn from a multivariate gaussian for each trial. We then use these new parameters to derive the new luminosity of the cluster within the spectral fit radius after zeroing out all background components. This procedure adequately accounts for the error in the complete spectral model rather than in just a simple model considering only cluster emission. We also include the errors in our surface brightness fit in our Monte Carlo simulation when we compute the total X-ray luminosity within r$_{200}.$ We carry out a total of 200 trials and tabulate our results in Table \ref{sourcelist}.

\section{Results}

\subsection{Cluster Morphology and Profiles}

The X-ray cluster emission from our sample clusters is generally smooth, mildly elliptical, and centered on the BCG. In Figure \ref{contourfig2}, we show the $0.3-8.0$ keV X-ray contours overlaid over the corresponding Digital Sky Survey\footnote[5]{The Digitized Sky Surveys were produced at the Space Telescope Science Institute under U.S. Government grant NAG W-2166. The images of these surveys are based on photographic data obtained using the Oschin Schmidt Telescope on Palomar Mountain and the UK Schmidt Telescope. The plates were processed into the present compressed digital form with the permission of these institutions.} (DSS) field. Abell 3693 has a clump to the southeast, although the central cluster appears relaxed. Abell 2877 and 3705 are exceptions that do not appear to be dynamically relaxed.   Abell 2877's diffuse emission is not centered on the BCG. Moreover, Abell 2877's low signal-to-noise and a bright, nearly central (2 arcmin from the cluster center) point source complicates the analysis. The NED database indicates that A2877 consists of two superposed clumps, which further suggests that it is unrelaxed. Abell 3705 has asymmetric, unsmooth contours and a clump to the northwest, and it is our sole BM Type III cluster. Because we model our clusters as relaxed and symmetric objects, the results for these two clusters are suspect. 

In the left panel of Figure \ref{sbspecplot1}, we plot the 1D surface brightness profiles with the point sources and particle background removed. Error bars represent the standard deviation of the mean surface brightness value for a given bin. Our results from fitting the single 2D beta model fits are in Table \ref{betafit}, and the radially averaged surface brightness fit residual profile (in units of $\chi$) is presented in a sub-panel beneath each cluster's surface brightness profile. We also compute 1D $\chi^2$ values using our best fit 2D model, which serve as an additional measure of the goodness-of-fit. Our fits to Abell 2877 and 3705 yield unphysical $\beta$ and $r_X$. We do not quote errors for Abell 2877 fit parameters because \emph{SHERPA}'s Monte Carlo error estimator fails to converge on a result. 
Because of the evidence suggesting that Abell 2877 and 3705 are unrelaxed, we do not include either in our final BCG+ICS versus cluster galaxy enrichment comparisons.

\subsection{Cluster Spectral Fits}
We plot the results of our spatially-resolved spectral fitting in Figure \ref{sbspecplot1} with temperature profiles in the middle panel and chemical abundance profiles in the right panel. Error bars denote $1\sigma$ values derived using \emph{XSPEC}. Several of our clusters have cool cores and steeply declining abundance gradients that peak at $\sim \textrm{Z}_{Fe,\odot}$ and fall to the canonical value of  0.3 Z$_{Fe,\odot}$ at large radii. These results illustrate the importance of spatially-resolved spectroscopy out
to large radii. Only four of our clusters, Abell 1651, Abell 2811, Abell 3693, and Abell 3705, do not show steep abundance gradients. Although the conventional wisdom is that clusters with isothermal profiles typically do not have abundance gradients \citep{degrandi}, we see here that even clusters that appear isothermal, Abell 2984, S84, and S540, have abundance gradients. In addition to abundance gradients, a few of our clusters show declining temperatures at large radii. This may be an artifact of the spectral fitting arising from an over-subtracted background, but similar declines are observed by \citet{vikhlinin} in their Chandra study of multiple clusters using a different analysis, and by \cite{snowden07}, which suggests a physical effect.

\subsection{Intracluster Gas and Fe Masses}

We use the following prescription to calculate the deprojected gas mass within any specified radius, which we determine by integrating the volume gas density profile (Eq. \ref{gasbeta}). We choose to evaluate quantities to a radius of r$_{500},$ the radius at which the cluster mass density is 500 times the Universe's current mass density and the largest radius for which 
gas and stellar mass measurements are available for most clusters. Reaching this radius often requires some extrapolation, so we also quote gas mass values out to the maximum radius of our spectral fits, r$_{spec}.$

Using the volume gas density profile, we compute the unprojected hydrogen gas mass enclosed within each annulus used in our spectral fits, multiply it by the spectrally-derived abundance to obtain Fe mass, and then use the solar photospheric Fe/H mass fraction \citep[0.0026;][]{anders89} to determine the physical quantity of Fe in solar masses present in that radial bin. For r $> $ r$_{spec}$, we assume a constant abundance equal to that at r$_{spec}.$ The recently revised solar values \citep{asplund} do not affect our results because our spectral fits also use the \citet{anders89} solar photospheric abundances and the derived Fe abundances should thus scale linearly with the difference between the two solar abundance catalogues.  We present the calculated hydrogen and Fe masses in Tables \ref{gasmass}, \ref{ferspec}, and \ref{fe200}. Typically, the extrapolation is less of an issue for gas masses than for Fe masses because the surface brightness fits that constrain the shape of the gas density extend farther than r$_{spec}$. Gas masses provided in these Tables are for hydrogen (H) because abundances are quoted relative to H by convention. To obtain the total gas mass, one must account for the mass fraction of helium and multiply the hydrogen gas mass by 1.33 (assuming primordial relative abundance). This adjustment is done in $\S$\ref{otherworks} for the comparison of our gas mass estimates with those presented in other studies.

\subsection{Systematic Error arising from Spherical Symmetry Assumption}

One possible source of systematic errors that could affect both our and previous studies is the assumption of spherical symmetry. Even relaxed looking clusters are often 
elliptical (Figure \ref{contourfig2}). Because the gas mass prediction is critically linked to the assumed geometry of the system, we test how the spherical assumption affects our results by carrying out the analysis on gas distributed in different ellipsoidal geometries. We start with the model gas density prescription outlined with $\beta = 2/3$ and $r_X = 5$ in $\S$\ref{sbfitting}, and  recast it into an elliptical form:
\begin{equation}
\rho_g(x,y,z) \propto \left[1+\frac{1}{r^2_X}\left[\left(\frac{x}{\epsilon_x}\right)^2 + \left(\frac{y}{\epsilon_y}\right)^2 + \left(\frac{z}{\epsilon_z}\right)^2 \right]\right]^{-3\beta/2},
\end{equation}
where the $\epsilon$'s are the ellipticities along each of the three principal axes. We calculate the gas density on a $401^3$ grid with the line-of-sight along the z-axis. We generate five different models: a spherical one, two oblate spheroids with ellipticities $(\epsilon_x = 1, \epsilon_y = 0.7, \epsilon_z = 1)$ and $(1,1,0.7),$ respectively, and two prolate spheroids with ellipticities $(1,0.7,0.7)$ and $(0.7,0.7,1),$ respectively. We also bound all of the distributions such that $((x-x_{cen})/\epsilon_x)^2+((y-y_{cen})/\epsilon_y)^2+((z-z_{cen})/\epsilon_z)^2 < r^2_{max}$ where $r_{max} = 200.$ Because the X-ray emission is thermal bremsstrahlung, we compute a surface brightness map by assuming isothermality, squaring the density profile, and integrating along the line of sight. We generate radial profiles by azimuthally averaging the surface brightness map.

Figure \ref{sphereerror} shows that the spherical model has the highest surface brightness, followed by the oblate and then prolate cases. By normalizing the profiles at a given radius, we see similarities in the projected profiles despite the different geometries; for example, all have similar slopes after the knee in the curve, which means that they have similar $\beta$ values. Starting with a cluster profile as shown in the right panel of Figure \ref{sphereerror}, we compute the gas mass for each ellipsoidal geometry and compare to the spherical case. We determine the upper and lower error bounds for the oblate and prolate cases. For the oblate case, the error ranges from +19.5\% to +11.5\%, relative to the spherical case, where the largest error corresponds to the cluster being flattened along the line of sight. For the prolate case, the upper and lower bounds are +17.5\% and +25.9\%, where the worse case error arises when the cluster is flattened along the line-of-sight. 
We conclude that, except for extreme geometries, the error arising from our spherical assumption is likely to be $\sim$ 15\%. 

\subsection{Comparison with Previous Work}
\label{otherworks}
As a check on our procedures, we compare our temperature, luminosity and gas mass measurements to those from previous studies. We first compare our luminosity measurements with the REFLEX \citep{bohringer04} values for the systems that we have in common. To do so, we recompute our $0.1-2.4$ keV luminosity within the REFLEX aperture. In the left panel of Figure \ref{rcomp}, we show the fractional difference between our respective luminosity measurements. In general, we agree to within 20\% and there is no systematic trend with cluster velocity dispersion. There are a few significant outliers, such as Abell 2984 and 3693, for which our luminosities are lower by as much as 60\%. There is no clear explanation for this discrepancy. We also compare the peak temperatures for seven of our clusters with those found in the BAX database \citep{sadat04} in the right panel of Figure \ref{rcomp}. In general, the values are consistent. Our values are systematically higher for four clusters (three of which have cool-cores), but this shift is likely to arise from our use of peak temperatures instead of emission-weighted temperatures, which are usually dominated by cool-cores. The one exception is Abell 2811, which has an isothermal temperature profile.

We also compare our spectral fit values for the three clusters, Abell 496, 3112, and 4059, that are in the recent cluster catalogue released by \cite{snowden07} with temperature, abundance, and surface brightness profiles. In all cases, our profiles are consistent with those derived by \cite{snowden07}. We do see a small systematic offset in the temperature profile of Abell 3112, but the difference is less than 10\%. In general, \cite{snowden07} measure temperatures to larger radii, but they measure abundance values to similar radii as we do. We do not attempt to extend our temperature fits to larger radii as that is not the focus of this paper.

Lastly, we carry out gas mass comparisons, which reflect on our ability to ultimately determine Fe masses. We search the literature for gas mass measurements within $r_{500}$ for our clusters \citep{reiprich01,castillo03,piffaretti05} and tabulate the results of our comparisons in 
Table \ref{comparison1}. 
For the comparison, we determine the gas mass interior to the $r_{500}$ defined in each comparison study and convert to their cosmologies. We hereafter refer to Einstein-de Sitter cosmology  ($\Omega_m = 1,$ $\Omega_\Lambda = 0$, $H_0 = 50$ km s$^{-1}$ Mpc$^{-1}$) as SCDM50. The methods used in papers with ROSAT data \citep{reiprich01,castillo03} involve fitting a projected gas mass profile to the observed surface brightness distribution, with slight variations in the choice of fitting function, although use of the beta profile is common. Our gas masses are consistent to within our uncertainties for one of three clusters, Abell 3112, from \citet{castillo03}. For Abell 496 and 1651, the results differ by 18\% and 11\%, respectively, but \cite{castillo03} do not provide error estimates for their gas masses so 
we cannot judge the significance of this disagreement. 
Our results agree with those of \citet{reiprich01} for five clusters.  

We also compare our gas mass values with those of \citet{piffaretti05}, who, like us, use XMM-Newton observations. Our masses are systematically higher than theirs by as much as 50\% for the same LCDM70 cosmology. They employ a more sophisticated spectral deprojection technique to derive their masses, yet our application of a deprojection technique to one of our clusters suggests that this makes only a modest change and is not responsible for the difference. Instead, the discrepancy arises from the gas density profiles. Our gas densities, independent of radius, are consistently higher than theirs by a factor of $\sim 1.6$. 

If the \citet{piffaretti05} results are closer to the correct value, which would suggest that all previous ROSAT measurements are incorrect, we are overpredicting the cluster Fe mass by $40-50$\% and hence overstating any discrepancy between the predicted Fe mass created by the ICS, BCG, and cluster galaxies and the measured Fe mass in the ICM ($\S$\ref{metalssec}).
Therefore, we carry out an additional test by plotting the relation between seven other clusters
that have gas masses measured by \cite{vikhlinin06} and velocity dispersions from
the literature \citep[see][]{gonzalez07}. We then compare this relation
to ours (Figure \ref{gascomparisons}).
We do not see a systematic offset between the mean trends found in our data and the \citet{vikhlinin06} data, suggesting further
that our gas masses are accurate.

\subsection{X-ray Scaling Relations}
As a further test of our spectral fitting techniques and the accuracy of the velocity dispersions of our clusters, we compare our sample to known $L_X-T-\sigma-$optical relations. We define the cluster temperature, which is a reflection of its dynamical mass, to be the highest temperature derived in our spatially-resolved spectral fits. This choice is reasonable because the derived temperatures are annularly-averaged and not likely to be affected by localized hot spots. In Figure \ref{xraydiag}, we show that the majority of our clusters fall within the errors of the $T-\sigma$ relation given by \citet{wu}: 
\begin{equation}
\sigma = 10^{2.49\pm0.03}T^{0.64\pm0.02},
\end{equation}
which is derived from a fit to 92 clusters. Outliers from this, and the other scaling
relations discussed below, are described individually in the Appendix.

As a further sanity check, we determine the $L_X-\sigma$ and $L_X-T$ relations for the 10 of our clusters with measured luminosities. We can adequately fit a power law in both cases (Figure \ref{xraydiag}), obtaining the following two relationships:
\begin{equation}
L_X=10^{36.11\pm2.17}\sigma^{2.83\pm0.70}
\end{equation}
and
\begin{equation}
L_X=10^{42.38\pm0.46}T^{2.79\pm0.77},
\end{equation}
where $L_X$ is the luminosity measured in the $0.5-2.0$ keV range. 
These relationships establish the validity 
of the X-ray reduction and analysis techniques employed here.

\section{Discussion}


There have been numerous previous attempts to explain the metals in the intracluster medium (ICM)  \citep[see][and references therein]{portinari04}. If cluster galaxies are the sole polluters of the ICM, which is the basic supposition of these studies, then it is not possible to produce the observed Fe mass using standard assumptions for the IMF, SN Ia rate as a function of time, and metal loss efficiency of galaxies \citep{portinari04}. 
This has led to arguments for a top-heavy or a variable IMF \citep{matteucci95,gibson97,moretti03,portinari04,loewenstein06}, which would reduce the metal mass locked in low-mass stars with lifetimes longer than a Hubble time (the locked-up fraction) and produce a larger number of Type II supernovae. These studies often assume that strong winds from the supernovae blow a significant fraction of the metals out of the galaxies, i.e., 
up to a 75\%  metal loss fraction from elliptical galaxies, which leaves behind the 25\% locked up in lower mass stars. Recent work produces a more consistent picture of the ICM enrichment through hydrodynamical simulations that rely on galactic superwinds \citep{romeo06} and possibly ram-pressure stripping \citep{kapherer07} in order to explain the observed trend in Fe abundance with redshift \citep{calura07}. Although these various models produce consistent results, the underlying prescriptions have yet to be observationally proven and/or calibrated. 

Often neglected in these analyses is the contribution from intracluster stars that pollute the ICM {\it in-situ}.
With the measurements of the cluster Fe mass and the luminosity of ICS in hand, we are in position to address a variety of questions
regarding the role of intracluster stars in the chemical enrichment history of clusters. 
Do intracluster stars play a significant role? If so, how does their contribution compare to 
that of stars in galaxies? Can the two components
together account for the metals observed in the ICM? Can we use these
constraints to place limits on the metal
mass loss from galaxies? Do we require non-standard assumptions to explain the metals?

\subsection{Metals from Intracluster Stars}
\label{metalssec}
We begin the calculation of the metal contribution from intracluster stars
using the parameters for the intracluster light measured by \citet{gonzalez05} and \citet{kelson}. We then calculate the mass-to-light ratio, corresponding SN rate, and chemical deposition rate using the PEGASE.2 evolutionary models \citep{fioc97} for 1 M$_\odot$ initial parcel of gas.
Scaling those results by the current mass of intracluster stars and integrating the SN rate over the lifetime of the stellar population, we calculate the Fe mass 
produced by the ICS. Because decomposition of the light from the brightest cluster galaxy and the intracluster stars is ambiguous, particularly in the \cite{kelson} data, which do not extend as far in radius as the \cite{gonzalez05} data, we compute the Fe mass generated by the combination of the BCG and intracluster stars and refer to that as the BCG+ICS contribution. Where we can decompose these two components because of abrupt changes in position angle or ellipticity, 
the BCG generally contributes only 10-20\% of the light \citep{gonzalez05}. Therefore, even though we present
results for the BCG and intracluster stellar components together, they do primarily reflect the metal contribution from intracluster stars. 

We make two important simplifying assumptions. First, we consider the ICS to be an old population originating from an initial, short-lived episode of star formation. 
\cite{williams07}, in their study of intracluster stars in Virgo, conclude that the majority of the ICS are old ($\gtrsim 10$ Gyr). Second, we assume that any
metals produced by the ICS find their way into the ICM, regardless of the origin of the ICS. Recent simulations suggest that the ICS are predominantly stripped from parent galaxies through tidal interactions \citep{murante07}. There are also observational claims that these stars 
form in gas being stripped from infalling cluster spirals \citep{sun07}. All the metals produced by 
stars that eventually become ICS, either before, during, or after the tidal interactions, are assumed to lie in the ICM. If, for some reason, the interactions that created the ICS were less efficient at removing metals from galaxies and if much of the metal enrichment due to these stars happened prior to their
becoming ICS, then we will overpredict the ICS Fe contribution. The second assumption can be tested observationally by measuring present-day intracluster and cluster galaxy SN Ia rates. There are indications that $\sim 20\%$ of cluster SN Ia explosions occur outside of galaxies \citep{galyam03, domainko04}, though these results are encumbered by large uncertainties. An extensive cluster SN Ia survey is underway to improve the precision of this result \citep{sand08}.

We model the ICS as a 14 Gyr old stellar system with an initial, single, 1 Gyr long, constant star formation rate burst and a standard Miller-Scalo initial mass function \citep{millerscalo}. We choose default PEGASE parameters: Model B yields for SN II ejecta \citep{woosley95}, no gas infall, no galactic winds, no nebular emission, and no extinction. We use 250 Myr time steps for the calculation. Using shorter, 1 Myr, time steps changes the final values only by a few percent. We set the initial metallicity to 0.004 by mass fraction. This is a plausible upper limit (0.2 Z$_\odot$), which is $\sim 2\times$ larger than that
of the old populations in local galaxies (e.g., the SMC \citep{harris04} and LMC \citep{harris08}). Taking a more conservative initial value of 0.0001 leads to a modest decline in Fe of 5\% (largely attributable to the change in mass-to-light (M/L) ratio), which is well within the other uncertainties. We do not use the evolving metal abundances calculated by PEGASE in any way other than indirectly through
the calculated M/L.   Hereafter, we refer to this model as our {\sl baseline model}.

The true nature of SN Ia, which are the dominant source of Fe, and their progenitors remains a subject of debate \citep{maoz08}. Currently, two possible pathways are theorized to produce Type Ia supernovae. In the single-degenerate (SD) case, a single white dwarf (WD) accretes matter from a close companion until it explodes upon reaching the Chandrasekar limit \citep{whelan73}. The double-degenerate (DD) model involves the merger of two binary WDs \citep{iben84} that leads to the explosion.
We model the SN Ia rate as a function of time by using the rates calculated by PEGASE, which uses a popular SD prescription \citep{greggio83}. This prescription has a free parameter, A,  the close binary fraction, that is manually tuned to match the present day SN Ia rate. We choose A $=$ 0.175 to match the observed present-day, cluster early-type SN Ia rate ($0.066^{+0.027}_{-0.020}$ SNuM, \citet{mannucci07}). The steep evolution in our model rate as a function of look-back time is also consistent with all other cluster SN Ia rate measurements carried out at different epochs \citep[$z \la 1$;][]{galyam02,sharon07}. The purpose of this modeling is to provide the SN rate over the lifetime of the stellar population. However, the dependences of the SN rate on environment and look-back time are poorly known. Adding to the uncertainties surrounding the SN rates, \cite{sb06} recently presented evidence for 
two classes of Ia's, prompt and delayed, whose relative importance will naturally vary over time and with environment.  We consider a crude empirical model incorporating these two SNe Ia classes at the end of this section. 

We adopt empirically determined Fe yields and define yield as the mass of Fe ejected into the interstellar medium per SN explosion. We use a Type Ia Fe yield of 0.7 $M_\odot$ \citep{maoz}, which is consistent with both the modeling of bolometric light curves of SNIa \citep{contardo00} and theoretical SN Ia yield calculations \citep{thielemann86}. Given the large scatter in Type II yields (0.0016 to 0.26 M$_\odot$; \cite{hamuy03}), we adopt an intermediate mean yield of 0.05 $M_\odot$ from \cite{elmhamdi}, who estimate the mass of $^{56}$Ni (which eventually decays to Fe) produced in a Type II SN by matching calculations to measured light curves and H$\alpha$ luminosities. We make the instantaneous recycling approximation by adopting a constant yield per Type II SN.

To scale the PEGASE results to the corresponding stellar mass of the ICS, we deproject the optical luminosity profile using a 3-D Hernquist profile \citep{hernquist90}, which is known to fit a deVaucouleurs profile in projected space. This procedure enables us to obtain the total optical light 
within spherical shells defined by the corresponding X-ray spectral fit annuli. The optical luminosity is converted to a stellar mass using an $i$-band or R-band mass-to-light ratio of 2.25 or 2.95, respectively. 
The enclosed mass is given as
\begin{equation}
M(r) = M_{tot}\frac{r^2}{(r+a)^2},
\end{equation}
where $M_{tot}$ is the total mass derived from the projected surface brightness profile
and $a$ is related to the scale length, $R_e$, through the relation $R_e \approx 1.8153a$.

First, we use the PEGASE model to calculate the Fe mass produced by the BCG+ICS and compare those values
to the observed ICM Fe masses. We do not consider the locked-up Fe fraction in the ICS because recent metallicity measurements of the ICS in Virgo give low values \citep[$\sim 0.1$ Z$_\odot$ on average;][]{williams07}. Consequently, only a negligible fraction of Fe ($\sim 2\%$) produced by the ICS is locked up in stars and virtually \emph{all} of the metals produced by the ICS go into the ICM.
Our results for volumes inside radii of r$_{spec}$ and r$_{500}$ are presented in Tables \ref{ferspec} and \ref{fe200}, respectively.  Column 2, in both 
Tables, contains the measured Fe mass within the specified radius, while Columns 3, 4, and 5 contain the predicted Fe mass contributed by the BCG only, the intracluster stars only, and the combination of these two components, respectively. We are
not able to decompose the BCG and ICS in all systems. The final column lists the fraction of the observed Fe mass contributed by the BCG+intracluster stars for the baseline  model. To estimate the uncertainties in the fractional contributions, we adopt 
a 15\% systematic error in the BCG+ICS Fe contribution arising from variations in M/L and from projection effects.  We incorporate the random error from X-ray Fe mass measurements. We do not include the uncertainty from the SN Ia rate in the individual cluster values, but include it in the averages
that we discuss below. We present the r$_{spec}$ results for the sake of completeness but do not use them in our work. 

We place all clusters on a common spatial scale by calculating all of the relevant quantities out to r$_{500}$. Because the ICS are more centrally distributed than the metals, this choice of radius is important. We chose r$_{500}$ because it requires the least amount of extrapolation from $\textrm{r}_{spec}$ (see Table \ref{gasmass}) and includes a significant portion of the 
cluster metals (cf. \cite{ettori05}, \cite{calura07}).  All of our clusters' Fe profiles are either flat or decreasing with radius (Figure \ref{sbspecplot1}). Therefore, our method of extrapolation will at worst overestimate the amount of Fe present within r$_{500}$ in clusters and thereby lead us to understate the contribution from intracluster stars within that radius. 

Our results indicate that intracluster stars do play a significant role in the enrichment history of clusters. In our baseline model,
the BCG+ICS contribute an average of $31^{+11}_{-9}$\% of the observed intracluster Fe within r$_{500}$. 
We show in Figure \ref{finalresult} (top panel) the fractional contribution of the BCG+ICS for each cluster with the dotted line representing the 31\% weighted average value. All averages are computed using $1/\textrm{error}^2$ as weights. The highly discrepant point with the lowest velocity dispersion is Abell 2984 and incidentally has the largest errors in fractional BCG+ICS contribution. The r$_{500}$ values for two lowest velocity dispersion clusters (Abell 2984 and Abell S 84) are only constrained by a single low velocity dispersion point in the calibration of the r$_{500}-\sigma$ relation \citep{gonzalez07}, and we suggest some caution in interpreting the results from these two 
clusters.
Because the BCG typically contributes much less Fe than the ICS in clusters where we can separate the BCG and ICS, we conclude that
the ICS are significant polluters of the intracluster medium within r$_{500}$ and need to be included in all future ICM metal budget accounting.

Next, we explore the potential effect of having two classes of SN Ia's. We 
restrict ourselves to the parameters of the two populations as specified by \cite{sb06}. 
Their model consists of a prompt population (with explosion delay time $\tau \sim$ 0.5 Gyr)
that scales in number with the star formation rate and of a delayed population ($\tau \sim$ few Gyr) 
that scales with the total stellar mass. More luminous (prompt) SNe Ia are observed preferentially in star-forming 
galaxies, while fainter (delayed) ones are seen in red galaxies with little star formation \citep{sb06}. 
Using the prescription suggested by \citet{sb06}, we derive the SN Ia rate using the star formation rates and stellar masses output by PEGASE for a 14 Gyr population. Most of the Fe enrichment is dominated by the prompt component at early times when the ICS progenitors are actively forming stars.
We obtain a BCG+ICS average fractional Fe contribution of {\bf $22^{+9}_{-9}$\%} within r$_{500}$, which is consistent with our one-component, baseline model.  
In contrast with the baseline model, the two-component model is relatively insensitive to the present day SN Ia rate (rescaling this delayed population upward by a factor of 3 increases the fractional
Fe contribution by $< 5$\%, because the prompt population is the dominant Fe contributor).
The \cite{sb06} prescription for the prompt rate is based on measurements of field galaxies
over a range of redshifts.  Because we do not know how the prompt component scales
with environment, we retain their scaling.
Although the baseline and two-component SN Ia models have different assumptions, they produce ICS Fe fractions in rough agreement. We do not explore more sophisticated theoretical models that vary the delay-time distributions (DTD) of SN Ia \citep[see][]{greggio05} to explain the Fe content in clusters \citep[e.g.][]{maoz,ettori05} because of large uncertainties in the explosion mechanism and properties of SN Ia progenitors.

Both the baseline and two-component SN Ia models predict that the ICS Fe contribution falls short of producing the observed Fe mass within the ICM. 
In some combination, we expect that additional enrichment channels (such as 
metal-injecting winds from galaxies)
and environment- and/or time-dependent SN Ia rates and yields will play roles in the final resolution of
the question of ICM enrichment.
We discuss metal loss from galaxies in \S \ref{galaxies}, and focus here on whether our
data independently support the claim of environment-dependent SN Ia rates \citep{mannucci07}.
(Our sample lies at one redshift, so assessing time-dependencies 
from it is not possible.)

One potential piece of evidence supporting the effect of environment 
is the trend between the model Fe fraction and $\sigma$  shown in Figure \ref{finalresult} (top panel).
A Spearman rank test demonstrates that the null hypothesis (no trend) can be excluded with $>$ 90\% confidence.
If the ICS contribute a {\sl fixed fraction} of the ICM Fe for all of our groups and clusters, then 
stars in more massive environments must eject more Fe per stellar mass. If we attribute these
Fe production differences entirely to differences in SN Ia rates, then we infer a
factor of roughly three increase in that rate from the low-$\sigma$ to high-$\sigma$ limits
of our range. 
This matches the factor of three increase between field and cluster SN Ia rates found by \cite{mannucci07}.
In this scenario, our baseline model  overpredicts the Fe fraction coming from the ICS
in group environments because we matched the present-day PEGASE SN Ia rate to that observed in clusters. In any case, the ICS alone cannot fully produce the ICM Fe.

Given the variation among measurements of field SNe rates discussed above, and the likely dependencies on environment and look-back time, the rates are a significant source of uncertainty in resolving the chemical enrichment problem in clusters.

\subsection{Metals from Galaxies}
\label{galaxies}
The previous section indicates that the ICS is an important, but not exclusive, contributor of Fe to the ICM. 
Therefore, we now estimate the contribution of galactic stars to the Fe ICM budget within r$_{500}$ for the seven \citet{gonzalez07} clusters for which we have optical photometry for all galaxies within r$_{500}$. We first subtract the BCG+ICS contribution calculated using the baseline model from the measured ICM Fe value. The remainder of the metals must come from galactic stars.  Once again we apply the baseline model (because the galaxy stellar population is dominated by old stars in early-types within r$_{500}$) to predict how much Fe galaxies produce. We then calculate the galactic metal loss efficiency such that the sum of the weighted fractional contribution of the BCG+ICS and galaxies is 1. On average, the galaxies need to lose $84^{+11}_{-14}$\% of the Fe that they produce for the sum of BCG+ICS and galaxies to account for 100\% of the observed Fe (Figure \ref{finalresult}; bottom panel). The choice of a single metal loss efficiency for all systems may not be appropriate as we over-predict the metals produced by two clusters, Abell 2984 and S 84. However, it is impossible to disentangle variations in SNIa rates (see above) and metal-loss efficiency with environment with the available data.

This result does not consider that some fraction of all  metals are locked up in the current stellar populations of these galaxies. If the galaxies within r$_{500}$ have an average metallicity that is close to solar \citep[see][]{portinari04}, approximately 25\% of the metals must be locked up in stars. 
Therefore, given the uncertainties cited above,
the baseline model satisfies two key constraints: 1) the combination of 
metals produced by the ICS and other cluster galaxies can account for the observed ICM Fe, and 2)
the inferred metal loss fraction from galaxies ($84^{+11}_{-14}$\%)
does not conflict with the locked-up metal fraction ($\sim$25\%).

This baseline model requires that cluster galaxies lose a substantial fraction of the metals that they produce.
While studies using sophisticated chemical models have shown that field early-type galaxies might
lose $\sim 80$\% of the metals (specifically Fe) they produce during their lifetime  
\citep{calura06}, other observational studies suggest a much lower metal loss fraction (measured from $\alpha$ elements) \citep[$\sim$ 35\%;][ who do not distinguish between galaxies in different environments]
{bouche07}. This large discrepancy can at least be partly explained by \cite{bouche07}'s focus on $\alpha$ element measurements and their use of the instantaneous recycling approximation (IRA) to compute galaxy metal loss efficiency. While the IRA may offer a reasonable approach for computing the production of $\alpha$ elements, it is not appropriate for computing the production of Fe because of the comparatively longer timescale for Fe production \citep{renzini93}. Fe is less likely to be locked-up in stars and more likely to be expelled into the IGM because the majority of Fe is produced after the epoch of star formation. Therefore, 35\% metal loss fraction serves as a lower bound for the amount of Fe that can be lost.

To examine the implications of the lower bound of the metal loss fraction on our model, 
we reverse the question and ask what the SN Ia rate needs to be
if the Fe loss fraction from galaxies is only 35\%.  Given that condition, 
increasing the SN Ia rate to roughly its $2\sigma$ upper-limit, {\it i.e.,} by a factor of
$1.8\times$, succeeds in generating all the ICM Fe.
Therefore, considering the large uncertainties in the SN Ia rates, we conclude that BCG+ICS+galaxy
models with galactic metal mass loss fractions ranging from $\sim 75$\% (the upper limit 
due to the locked-up fraction) down to 35\% can reproduce the ICM Fe abundances.

\section{Conclusions}
We present X-ray surface brightness, temperature, and Fe abundance profiles out to 
$\sim r_{500}$ ($\sim 0.6$ times r$_{200}$, the virial radius) for a set of twelve nearby ($z\lesssim 0.1$) clusters with extensive optical photometry to quantify the relative contributions of intracluster stars (ICS) and cluster galaxies to intracluster medium (ICM) metal enrichment.
We study systems with velocity dispersions of $500 < \sigma < 1000$ km s$^{-1}$, measuring peak X-ray temperatures and $0.5-2.0$ keV luminosities within r$_{500}$ of $2-6$ keV and $0.3 - 3.4 \times 10^{44}$ ergs s$^{-1},$ respectively. The $T-\sigma$, $L_X-\sigma$, and  $L_X-T$ scaling relations for our systems are generally consistent with 
expectations. Two-dimensional X-ray surface brightness profiles reveal substructure in most of our clusters ranging from spiral-like structures to subclumps to central asymmetries. 

The majority (9 of 12) of the clusters, even those with isothermal profiles,  show steep abundance gradients that approach 
the canonical abundance of $\sim$ 0.3 $\textrm{Z}_{Fe,\odot}$ at large radii. Four of these 9 have cool cores. The radial variation in abundance makes necessary spatially-resolved spectroscopy to accurately quantify the radial distribution of metals, which is in turn necessary for self-consistent chemical evolution modeling.

The stellar component consisting of the brightest cluster galaxy plus intracluster stars (BCG+ICS) contributes on average a non-negligible fraction, $31^{+11}_{-9}$\%, of 
the ICM's Fe within r$_{500}$ for our baseline chemical evolution model.
We also calculate the Fe yield for an alternate, 
two-component (prompt and delayed), empirical SN Ia model \citep{sb06}.
Even though the baseline and two-component SN Ia models
have different assumptions, they produce BCG+ICS average fractional Fe contributions that are similar:  
$31^{+11}_{-9}$\% versus {\bf $22^{+9}_{-9}$\%} within r$_{500}$, respectively.
For the seven clusters in which
we know the relative contribution of the BCG and ICS, the ICS contribute 80\% on average of the combined BCG+ICS Fe, indicating that the ICS significantly enrich the intracluster medium within r$_{500}$ and must be included in any enrichment model for the ICM.

Because the BCG+ICS component cannot account for all of the Fe in the ICM, we 
consider the combined
effect of BCG+ICS {\sl and} other cluster galaxies on ICM enrichment. 
We then find that we can account for all the Fe within r$_{500}$ and that the required galactic metal
loss fraction ($84^{+11}_{-14}$\%) does not conflict with the fraction of metals 
still locked up in galactic stars ($\sim$25\%).
While this metal loss efficiency is large, it is consistent with other 
estimates \citep[$\sim $80\%;][]{calura06}.  It is also worth noting that, given the large
uncertainties in the SN Ia rates, the required metal loss fraction might be significantly lower than this initial estimate. 
For example, if we increase the rate of SN Ia in cluster galaxies by a factor of $1.8$,
to its upper $2\sigma$ bound \citep{mannucci07}, 
we can produce all the ICM's Fe with a galactic metal loss fraction of only $\sim$35\%,
which is consistent with other metal loss fraction estimates \citep{bouche07}.
As a result, it is possible to
make a full accounting of the ICM's Fe --- without resorting to extreme assumptions about
the stellar initial mass function or pre-enrichment --- for a large range of plausible
galactic metal mass loss fractions (35 - 75\%, the upper limit due to the locked-up fraction). 
Reducing the uncertainties in the SN Ia rates, 
including measurements of the likely dependencies on redshift and galaxy environment, is the 
critical next step in understanding the enrichment history of the intracluster medium.

\acknowledgements
We would like to thank the following individuals: Steve Snowden and Kip Kuntz for our lengthy discussions on XMM data analysis and EPIC background subtraction, and for the XMM-ESAS code they provided us for flare filtering and background subtraction, Kathy Romer  for her contributions in the initial stages of the project, David Buote and John Mulchaey for several fruitful discussions about X-ray cluster data analysis, and Romeel Dave, Daniel Eisenstein, Phil Hinz, and Michael Meyer for helpful comments on 
earlier versions of this paper. We would also like to thank the anonymous referee whose comments were very helpful. DZ acknowledges financial support for this work from a Guggenheim fellowship, NASA LTSA award NNG05GE82G, and NSF grant AST-0307482. AIZ acknowledges financial support from NASA LTSA award NAG5-11108 and from NSF grant AST-0206084. DZ and AIZ also want to thank KITP for its hospitality and financial support through NSF grant PHY99-07979 and the NYU Physics department and the Center for Cosmology and Particle Physics for their generous support and hospitality during their sabbatical there.  This research has made use of the X-ray Clusters Database (BAX) which is operated by the Laboratoire d'Astrophysique de Tarbes-Toulouse (LATT), under contract with the Centre National d'Etudes Spatiales (CNES), and the the NASA/IPAC Extragalactic Database (NED) which is operated by the Jet Propulsion Laboratory, California Institute of Technology, under contract with the National Aeronautics and Space Administration.

\clearpage

\begin{deluxetable}{lcccccccc}
\tablewidth{0pt}
\small
\tablecolumns{8}
\tablenum{1}
\tablecaption{Observed X-ray clusters and their properties \label{sourcelist}}
\tablehead{
\colhead{Name} & 
\colhead{$\langle z \rangle$} & 
\colhead{BM} & 
\colhead{$\sigma$} & 
\colhead{r$_{500}$ } & 
\colhead{r$_{200}$} &
\colhead{L$_{\mathrm{X}}$} & 
\colhead{T$_{\mathrm{peak}}$} \\
&
&
&
(km s$^{-1}$) &
(Mpc) &
(Mpc) &
(10$^{44}$ ergs s$^{-1}$) &
(keV)
}
\startdata
Abell 496 & 0.0329 & I &  $743^{+41}_{-39}$ & $1.033$ & $1.599$ & $1.81^{+0.06}_{-0.08}$& 4.39 $\pm$ 0.08 \\
Abell 1651\tablenotemark{*} & 0.0845 & I-II & $990^{+110}_{-100}$ & $1.397$ & $2.164$ & $3.37^{+0.16}_{-0.16}$ & 6.16 $\pm$ 0.22 \\
Abell 2811\tablenotemark{*}& 0.1079 & I-II & $860^{+110}_{-100}$ & $1.202$ & $1.862$ & $2.06^{+0.10}_{-0.11}$ & 5.76 $\pm$ 0.20\\
Abell 2877 &  0.0247 & I & $999^{+77}_{-71}$ & $1.406$ & $2.177$ & -- & 3.27 $\pm$ 0.14\\
Abell 2984\tablenotemark{*} &  0.1042 & I & $490^{+112}_{-91}$ & $0.658$ & $1.020$ & $0.33^{+0.02}_{-0.03}$ & 2.01 $\pm$ 0.05 \\
Abell 3112\tablenotemark{*} &  0.0750 & I & $940^{+140}_{-120}$ & $1.324$ & $2.050$ & $2.79^{+0.04}_{-0.04}$ & 5.24 $\pm$ 0.11 \\
Abell 3693\tablenotemark{*}& 0.1237 & - & $1030^{+150}_{-130}$ & $1.452$ & $2.249$ & $0.83^{+0.04}_{-0.07}$& 4.21 $\pm$ 0.14\\
Abell 3705\tablenotemark{*} & 0.0906 & III & $1010^{+80}_{-80}$ & $1.426$ & $2.209$ & -- & 3.71 $\pm$ 0.22\\
Abell 4010\tablenotemark{*}& 0.0963 & I-II & $630^{+150}_{-120}$ & $0.867$ & $1.343$ & $1.46^{+0.08}_{-0.14}$& 4.44 $\pm$ 0.34 \\
Abell 4059 &  0.0475 & I & $653^{+74}_{-67}$ & $0.901$ & $1.396$ & $1.23^{+0.06}_{-0.07}$& 4.14 $\pm$ 0.13 \\
Abell S 84\tablenotemark{*} & 0.1100 & I & $520^{+160}_{-120}$ & $0.705$ & $1.091$ & $0.76^{+0.13}_{-0.08}$& 4.28 $\pm$ 0.18 \\
Abell S540 & 0.0358 & I & $760^{+36}_{-35}$ & $1.057$ & $1.638$ & $0.33^{+0.01}_{-0.03}$& 2.65 $\pm$ 0.08 \\
\enddata
\tablecomments{BM = Bautz-Morgan type. Velocity dispersions are from \cite{zaritsky05} or calculated for this paper using the method outlined in \cite{zaritsky05} using radial velocities from the NASA Extragalactic Database. Luminosities are measured in the $0.5-2.0$ keV range out to r$_{200}.$ Our data do not extend that far in radius and therefore we extrapolate our surface brightness fits to obtain the enclosed X-ray luminosity. The quoted temperatures are the peak values obtained from our spectral fits. We choose the peak value as it best reflects the depth of the cluster potential, especially for clusters with a cool-core.}
\tablenotetext{*}{\citet{gonzalez05} clusters.}
\end{deluxetable}

\begin{deluxetable}{lccccccc}
\tablewidth{0pt}
\small
\tablenum{2}
\tablecolumns{6}
\tablecaption{X-ray data properties \label{dataproperty}}
\tablehead{
\colhead{Name} & 
\colhead{Obs ID\tablenotemark{1}} & 
\colhead{Date} &
\colhead{Type\tablenotemark{2}} & 
\colhead{Filter}&
\colhead{t$_{\mathrm{MOS}}$} &
\colhead{t$_{\mathrm{MOS,ff}}$} & 
\colhead{r$_{\mathrm{outer}}$} \\
& 
& 
& 
& 
& 
(ks) & 
(ks) & 
(arcmin)
}

\startdata
Abell 496   & 0135120201 & 2001 Feb 01 &XSA & Thin & 29.5 & 16.1 & 10 \\
Abell 1651 & 0203020101 & 2004 Jul 01& GO  & Thin & 15.2 & 8.0 & 6 \\
Abell 2811 & 0404520101 & 2006 Nov 28 & GO & Medium & 24.4 & 22.7 & 5\\
Abell 2877 & 0204540201 & 2004 Nov 23 &XSA & Thin & 21.9 & 20.1 & 10 \\
Abell 2984 & 0201900601 & 2004 Dec 27 &XSA & Thin & 29.2 & 27.2 & 4 \\
Abell 3112 & 0105660101 & 2000 Dec 24 &XSA & Medium & 23.3 & 22.4 & 6 \\
Abell 3693 & 0404520201 & 2006 Oct 14 & GO & Medium & 34.6 & 29.4 & 3\\
Abell 3705 & 0203020201 & 2004 Oct 06 & GO & Medium & 24.4 & 13.9 & 6 \\
Abell 4010 & 0404520501 & 2006 Nov 13 & GO & Thin & 19.6 & 18.5 & 6\\
Abell 4059 & 0109950101 & 2000 Nov 24 &XSA & Thin & 29.4 & 12.4 & 9 \\
Abell S 84 & 0201900401 & 2004 Dec 04 &XSA & Thin & 34.6 & 17.7 & 4 \\
Abell S540 & 0149420101 & 2002 Oct 11 &XSA & Medium & 18.0 & 11.1 & 8 \\
\enddata
\tablenotetext{1}{XMM-Newton observation identification.}
\tablenotetext{2}{XSA denotes data from the XMM-Newton Science Archive; GO denotes data from our Guest Observer programs.}

\end{deluxetable}

\begin{deluxetable}{lcccc}
\tablewidth{0pt}
\tablenum{3}
\tablecolumns{5}
\small
\tablecaption{Best fit 2D beta model parameters  \label{betafit}}
\tablehead{
\colhead{Name}  & 
\colhead{$\beta$} & 
\colhead{$r_\mathrm{X}$} & 
\colhead{$n_{H,0}$} & 
\colhead{$\chi^2_{\mathrm{DOF}}$\tablenotemark{1}} \\
&
&
(kpc) &
(cm$^{-3}$) &
}
\startdata
Abell 0496 & 0.473$^{+0.001}_{-0.001}$ & 21.1$^{+0.3}_{-0.3}$ & 0.033$^{+0.002}_{-0.002}$ & 6.6\\
Abell 1651 & 0.549$^{+0.006}_{-0.006}$ & 90.7$^{+2.5}_{-2.5}$ & 0.010$^{+0.005}_{-0.004}$ & 1.3 \\
Abell 2811 & 0.66$^{+0.01}_{-0.01}$ & 141.7$^{+3.7}_{-3.6}$ & 0.006$^{+0.003}_{-0.003}$ & 1.1\\
Abell 2877 & 0.28 & 0.38 & -- & 2.9 \\
Abell 2984 & 0.501$^{+0.009}_{-0.008}$ & 33.6$^{+2.2}_{-2.6}$ & 0.012$^{+0.001}_{-0.001}$ & 1.2\\
Abell 3112 & 0.564$^{+0.003}_{-0.003}$ & 51.6$^{+1.1}_{-1.2}$ & 0.020$^{+0.001}_{-0.001}$ & 1.7\\
Abell 3693 & 0.52$^{+0.01}_{-0.01}$ & 82.0$^{+5.5}_{-4.8}$ & 0.0052$^{+0.0006}_{-0.0005}$ & 1.7\\
Abell 3705 & -- & -- & -- & --\\
Abell 4010 & 0.500$^{+0.004}_{-0.003}$ & 24.2$^{+0.9}_{-0.9}$ & 0.031$^{+0.002}_{-0.002}$& 2.8\\
Abell 4059 & 0.497$^{+0.004}_{-0.004}$ & 43.1$^{+1.1}_{-1.0}$ & 0.013$^{+0.006}_{-0.005}$ & 1.1\\
Abell S  84 & 0.58$^{+0.01}_{-0.01}$ & 66.8$^{+3.5}_{-3.4}$ & 0.0078$^{+0.0007}_{-0.0006}$ & 1.3\\
Abell S540 & 0.49$^{+0.01}_{-0.01}$ & 39.4$^{+2.8}_{-3.1}$ &0.0074$^{+0.0009}_{-0.0007}$& 1.2\\
\enddata
\tablenotetext{1}{One-dimensional $\chi^2$ value normalized by the degrees of freedom (DOF).} 
\end{deluxetable}

\begin{deluxetable}{lcccc}
\tablewidth{0pt}
\tablenum{4}
\tablecolumns{6}
\small
\tablecaption{X-ray derived cluster hydrogen gas mass \label{gasmass}}
\tablehead{
\colhead{Name}  & 
\colhead{r$_{\mathrm{spec}}$} & 
\colhead{r$_{\mathrm{spec}}$/r$_{500}$} & 
\colhead{M$_{H,r_{spec}}$} & 
\colhead{M$_{H,r_{500}}$} \\
&
(kpc) &
&
(10$^{12}$ M$_\odot$) &
(10$^{13}$ M$_\odot$)
}
\startdata
Abell 0496 & 392.7 & 0.38 & 6.1$^{+0.1}_{-0.1}$ & 2.85$^{+0.14}_{-0.12}$ \\
Abell 1651 & 570.5 & 0.41 & 17.8$^{+0.2}_{-0.2}$ & 6.60$^{+0.16}_{-0.16}$ \\
Abell 2811 & 590.5 & 0.49 & 15.5$^{+0.2}_{-0.2}$ & 3.90$^{+0.01}_{-0.01}$\\
Abell 2984 & 458.9 & 0.70 & 4.4$^{+0.2}_{-0.2}$  & 0.77$^{+0.03}_{-0.04}$ \\
Abell 3112 & 511.9 & 0.39 & 12.08$^{+0.09}_{-0.09}$ & 4.45$^{+0.06}_{-0.07}$ \\
Abell 3693 & 399.7 & 0.28 & 5.0$^{+0.1}_{-0.1}$& 3.65$^{+0.18}_{-0.18}$\\
Abell 4010 & 642.0 & 0.74 & 12.2$^{+0.2}_{-0.2}$& 1.93$^{+0.04}_{-0.04}$\\
Abell 4059 & 501.9 & 0.56 & 8.4$^{+0.1}_{-0.1}$ &  2.09$^{+0.03}_{-0.03}$ \\
Abell S  84 & 481.0 & 0.68 & 6.0$^{+0.2}_{-0.2}$ &  1.01$^{+0.04}_{-0.05}$ \\
Abell S540 & 339.8 & 0.32 & 2.33$^{+0.05}_{-0.05}$ & 1.41$^{+0.07}_{-0.07}$ \\
\enddata
\end{deluxetable}

\begin{deluxetable}{lccccc}
\tablewidth{0pt}
\tablenum{5}
\tablecolumns{5}
\small
\tablecaption{Cluster gas mass comparison with literature \label{comparison1}}
\tablehead{
\colhead{Name} & \colhead{r$_{500}$} & \colhead{Ref M$_{g,r_{500}}$} & \colhead{Our M$_{g,r_{500}}$} & \colhead{Ref} & \colhead{Cosmology}\\
& (kpc) & ($10^{13}$ M$_\odot$) & ($10^{13}$ M$_\odot$) & &
}
\startdata
Abell 496	  & 1420$\pm 30$  & 6.9 & 8.4$^{+0.5}_{-0.4}$ & 1 & SCDM50\\
 & 1240$^{+20}_{-20}$ & 6.8$^{+0.2}_{-0.3}$ & 6.8$^{+0.3}_{-0.3}$& 2 & SCDM50\\
& 668$\pm 96$  & 1.2$\pm 0.3$ & 1.9$^{+0.5}_{-0.4}$ & 3 & LCDM70 \\
Abell 1651 & 1550$\pm 50$ &12.4 & 14.0$^{+0.6}_{-0.8}$& 1 & SCDM50\\
& 1730$^{+70}_{-80}$ & 15.1$^{+1.2}_{-1.2}$ & 16.4$^{+0.9}_{-1.2}$& 2 & SCDM50\\
Abell 3112 & 1440$^{+90}_{-140}$ & 9.4 & 9.4$^{+1.4}_{-0.7}$  & 1 & SCDM50 \\
& 1530$^{+110}_{-160}$ & 10.5$^{+1.3}_{-1.6}$ & 10.2$^{+1.6}_{-0.9}$ & 2 & SCDM50\\
& 1048$\pm 317$ & 2.8$\pm 0.6$ & 4.3$^{+2.1}_{-1.6}$ & 3 & LCDM70\\
Abell 4059 & 1400$^{+60}_{-60}$ & 6.4$^{+0.6}_{-0.5}$ & 7.4$^{+0.3}_{-0.6}$ & 2 & SCDM50\\
& 1247$\pm 304$ & 2.4$\pm 0.6$ & 4.6$^{+1.6}_{-1.7}$ & 3 & LDCM70 \\
Abell S540 & 1080$^{+140}_{-120}$ & 2.4$^{+0.8}_{-0.6}$ & 2.6$^{+0.5}_{-0.5}$ & 2 & SCDM50 \\
\enddata
\tablecomments{r$_{500}$ values taken from references.}
\tablerefs{(1) \citet{castillo03}; (2) \citet{reiprich01}; (3) \citet{piffaretti05}}
\end{deluxetable}

\begin{deluxetable}{lccccc}
\tablewidth{0pt}
\tablenum{6}
\tablecolumns{6}
\small
\tablecaption{Deprojected BCG and ICS Fe contribution out to $\mathrm{r}_{\mathrm{spec}}.$  \label{ferspec}}
\tablehead{
\colhead{Name}  &\colhead{M$_{Fe,\mathrm{r_{spec}}}$} & \colhead{M$_{Fe,BCG,\mathrm{r_{spec}}}$} & \colhead{M$_{Fe,ICS,\mathrm{r_{spec}}}$}  & \colhead{M$_{Fe,BCG+ICS,\mathrm{r_{spec}}}$} & \colhead{M$_{Fe,BCG+ICS}$/M$_{Fe,\mathrm{r_{spec}}}$}  
}
\startdata
Abell 0496 & 5.7$^{+0.4}_{-0.4}$ & -- & --  &  5.8 & 1.02$^{+0.17}_{-0.17}$ \\
Abell 1651 & 16.9$^{+2.8}_{-2.7}$ &  7.5 & 3.5 & 11.0 & 0.65$^{+0.14}_{-0.15}$\\
Abell 2811 & 15.1$^{+2.5}_{-2.4}$ & 1.8 & 8.5 & 10.2 & 0.68$^{+0.15}_{-0.15}$\\
Abell 2984 & 4.9$^{+0.8}_{-0.7}$ & 1.1 & 8.9 &  10.0 & 2.03$^{+0.42}_{-0.45}$\\
Abell 3112 & 12.1$^{+1.2}_{-1.1}$ & 1.0 & 10.6 & 11.6  & 0.96$^{+0.17}_{-0.17}$\\
Abell 3693 & 4.5$^{+1.1}_{-1.1}$ & 1.7 & 4.9 & 6.7 & 1.48$^{+0.43}_{-0.43}$ \\
Abell 4010 & 16.6$^{+2.8}_{-2.9}$ & 1.7 & 8.3 & 9.9 & 0.60$^{+0.14}_{-0.14}$ \\
Abell 4059 & 7.4$^{+1.0}_{-1.0}$ & -- & -- &  6.4 & 0.86$^{+0.17}_{-0.17}$\\
Abell S  84 & 7.5$^{+1.3}_{-1.4}$ &  0.4 & 7.7 & 8.1  & 1.08$^{+0.26}_{-0.25}$\\
Abell S540 & 2.4$^{+0.3}_{-0.3}$ & -- & -- &  6.6 & 2.77$^{+0.54}_{-0.54}$\\
\enddata
\tablecomments{All Fe masses are in $10^9 M_\odot$ units. These results are derived from our baseline PEGASE model \citep{fioc97}, which reproduces the present-day current cluster early-type SNIa rate \citep{mannucci07}. For the two-component SN Ia model values, divide the BCG and ICS Fe masses, and the Fe fractional contributions by $1.39.$}
\end{deluxetable}

\begin{deluxetable}{lccccc}
\tablewidth{0pt}
\tablenum{7}
\tablecolumns{6}
\small
\tablecaption{Deprojected BCG and ICS Fe contribution out to $\mathrm{r}_{500} $ \label{fe200}}
\tablehead{
\colhead{Name}  &\colhead{M$_{Fe,\mathrm{r_{500}}}$} & \colhead{M$_{Fe,BCG,\mathrm{r_{500}}}$ } & \colhead{M$_{Fe,ICS,\mathrm{r_{500}}}$}  & \colhead{M$_{Fe,BCG+ICS,\mathrm{r_{500}}}$} & \colhead{M$_{Fe,BCG+ICS}$/M$_{Fe,\mathrm{r_{500}}}$} }
\startdata
Abell 0496 & 23.3$^{+3.4}_{-3.0}$ & -- & --  &   6.5 & 0.29$^{+0.06}_{-0.06}$\\
Abell 1651 & 61.9$^{+13.6}_{-12.9}$ & 7.7 & 4.5 & 12.2 & 0.20$^{+0.05}_{-0.05}$\\
Abell 2811 & 39.0$^{+8.9}_{-8.5}$ & 1.8 & 9.5 & 11.3 & 0.29$^{+0.08}_{-0.08}$\\
Abell 2984 & 5.9$^{+1.0}_{-1.0}$ & 1.1 & 10.0 &  11.1 & 1.88$^{+0.43}_{-0.43}$\\
Abell 3112 & 36.0$^{+7.3}_{-6.7}$ & 1.0 & 12.1 & 13.1 & 0.36$^{+0.09}_{-0.09}$\\
Abell 3693 & 34.6$^{+12.9}_{-13.9}$ & 1.8 & 8.1 & 9.8 & 0.28$^{+0.12}_{-0.11}$ \\
Abell 4010 & 24.9$^{+4.9}_{-5.2}$ & 1.7 & 8.9 & 10.5 & 0.42$^{+0.11}_{-0.11}$\\
Abell 4059 & 14.7$^{+3.8}_{-3.8}$ & -- & -- &  6.9 & 0.47$^{+0.14}_{-0.14}$ \\
Abell S 84 & 12.3$^{+2.2}_{-2.5}$ &  0.4 & 8.1 & 8.4  & 0.69$^{+0.17}_{-0.16}$ \\
Abell S540 & 5.7$^{+2.3}_{-1.5}$ & -- & -- &  7.3 & 0.55$^{+0.15}_{-0.15}$\\
\enddata
\tablecomments{All Fe masses are in $10^9 M_\odot$ units. These results are derived from our baseline PEGASE model \citep{fioc97}, which reproduces the present-day current cluster early-type SNIa rate \citep{mannucci07}. For the two-component SN Ia model values, divide the BCG and ICS Fe masses, and the Fe fractional contributions by $1.39.$}
\end{deluxetable}

\clearpage
\begin{figure}
\epsscale{0.8}
\plotone{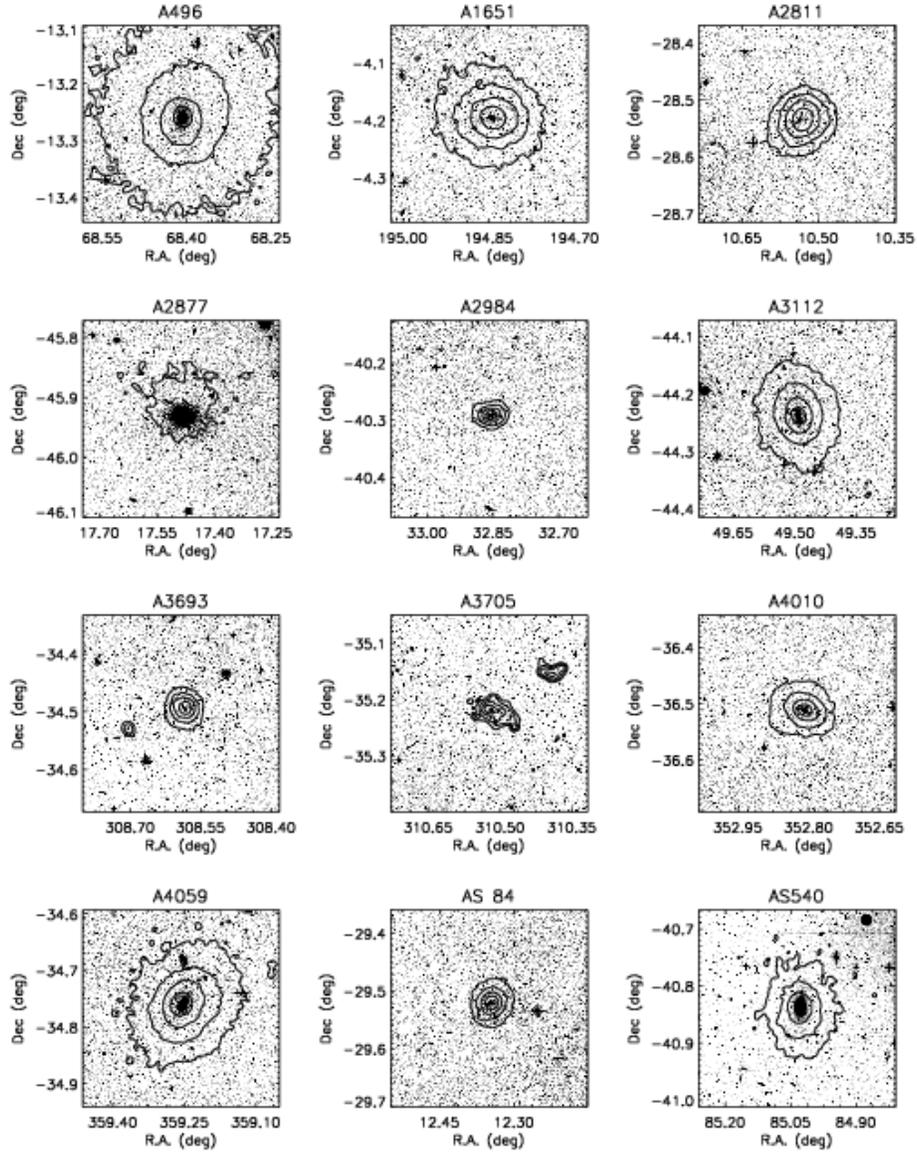}
\figurenum{1}
\caption{15\arcmin\ by 15\arcmin\ Digital Sky Survey (DSS) images with overlaid $0.3-8.0$ keV X-ray contours (particle background subtracted) for all observed clusters. The DSS images are stretched by first carrying out a 3$\sigma$ from the mean clip of all pixels. The lower and upper limits of the stretched image are chosen to be the mean of the clipped pixels and the mean plus 3$\sigma$, respectively. The point sources have been removed in the smoothed X-ray maps used to generate contours. Contour intervals are logarithmically spaced, spanning 5 intervals from the lowest value of 2$\times$10$^{-6}$ counts s$^{-1}$ arcsec$^{-2}$ to the peak value in the smoothed map. In general, the clusters are ``relaxed," i.e. are roughly spherical, and the contours are centered about the brightest cluster galaxy (BCG). Abell 2877 and 3705 are exceptions. The X-ray centroid of the diffuse emission of Abell 2877 is not centered on the BCG. Abell 3705, which is the only BM Type III cluster in our sample, is not spherical and may be in the process of a merger with the NW clump. 
\label{contourfig1} \label{contourfig2}}
\end{figure}

\clearpage
\begin{figure}
\centering
\epsscale{0.9}
\plotone{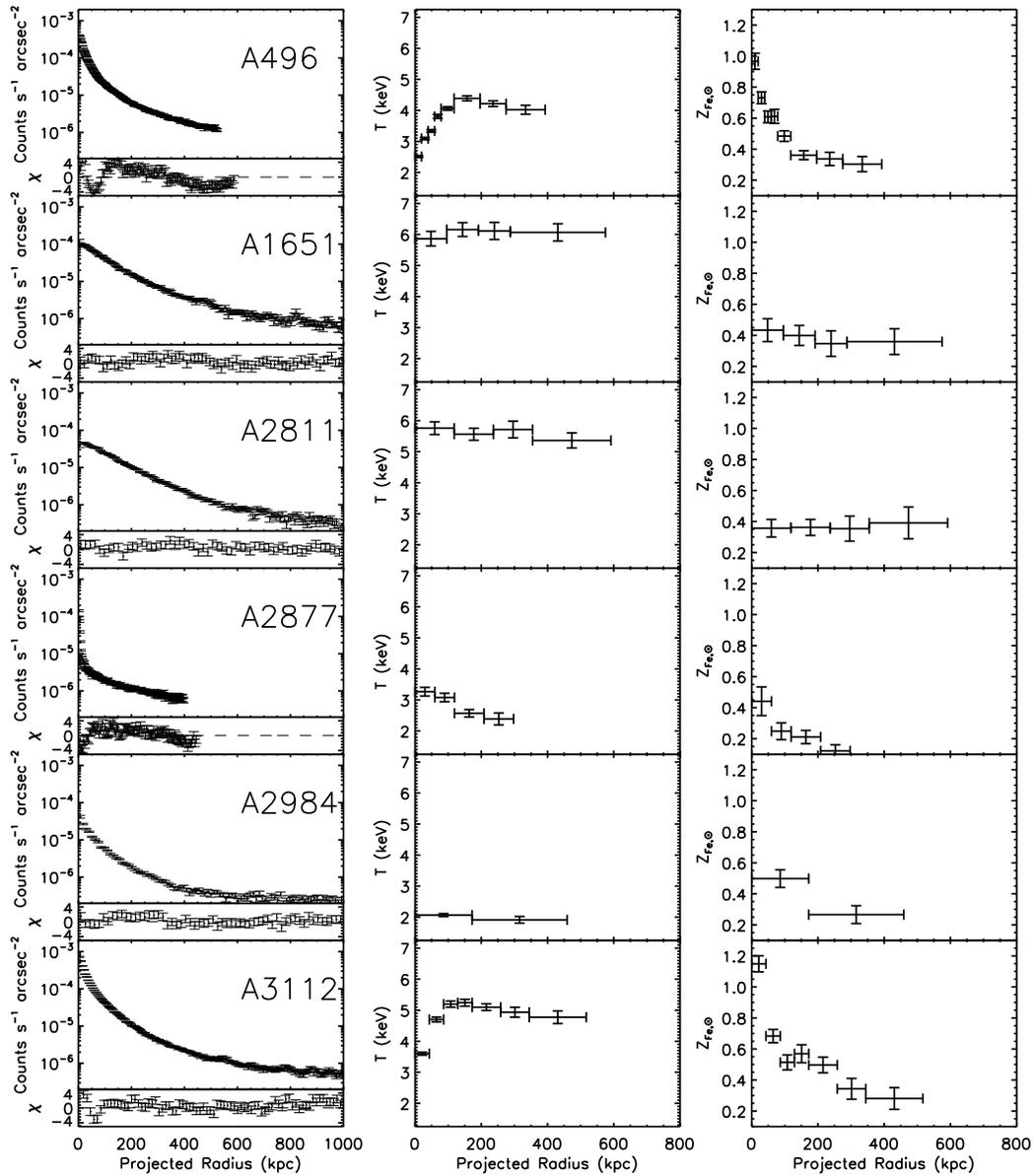}
\figurenum{2}
\caption{Surface brightness [$0.3-8.0$ keV] profiles and 1D fit residual in units of $\chi$ (left), temperature profiles (middle), and metallicity profiles (right) of the sample of clusters. The particle background has been subtracted from the surface brightness profiles, leaving only the cosmic background.  Several clusters with cool cores have temperature profiles that rise from the center, peak, and then drop
at larger radii. A majority of the clusters have radial abundance gradients that flatten to Z $\sim$ 0.3 Z$_{Fe,\odot}$ at large radii. The two clusters (Abell 1651 and 2811) with no obvious abundance gradients are also isothermal. Abell 3705 has such low signal-to-noise that its surface brightness fit fails to converge, as shown by its residual plot. 
\label{sbspecplot1}}
\end{figure}
\clearpage
\centering
{\plotone{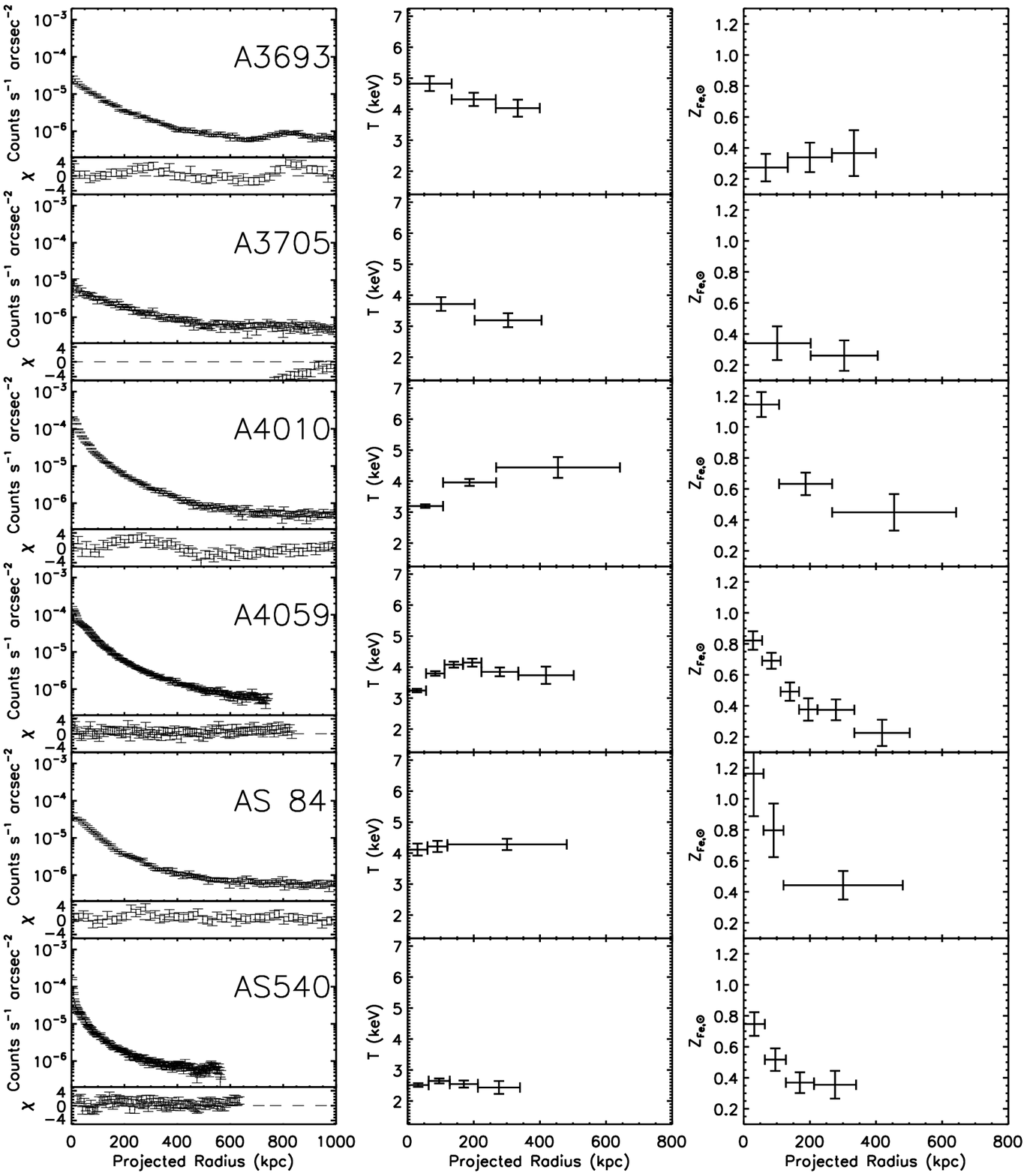}}\\
\centerline{Fig. 2. --- {Continued.}}
\clearpage

\begin{figure}
\centering
\epsscale{0.7}
\plotone{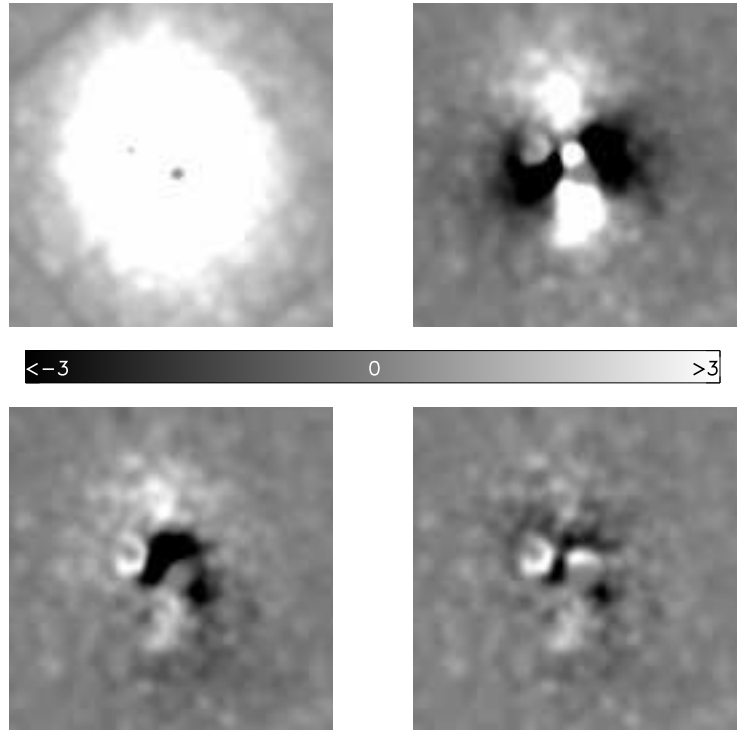}
\figurenum{3}
\caption{Abell 3112 2D surface brightness profile (10\arcmin \ on each side) fitting residuals. The data are smoothed by a 25\arcsec \ gaussian kernel to make the residuals easier to see. The color bar is in units of counts. The original data with the point sources removed are shown in the top left. This plot illustrates the ability of different types of models to fit the observed surface brightness. Spherically symmetric model fits (top right) result in asymmetric residuals because of the elliptical nature of cluster emission; a single component elliptical beta model (bottom left) fits fairly well, but only marginally better than the spherical fit. Adding an additional beta model (bottom right) to the fit again only improves the fit marginally. \label{2dresiduals}}
\end{figure}

\begin{figure}
\epsscale{0.8}
\plotone{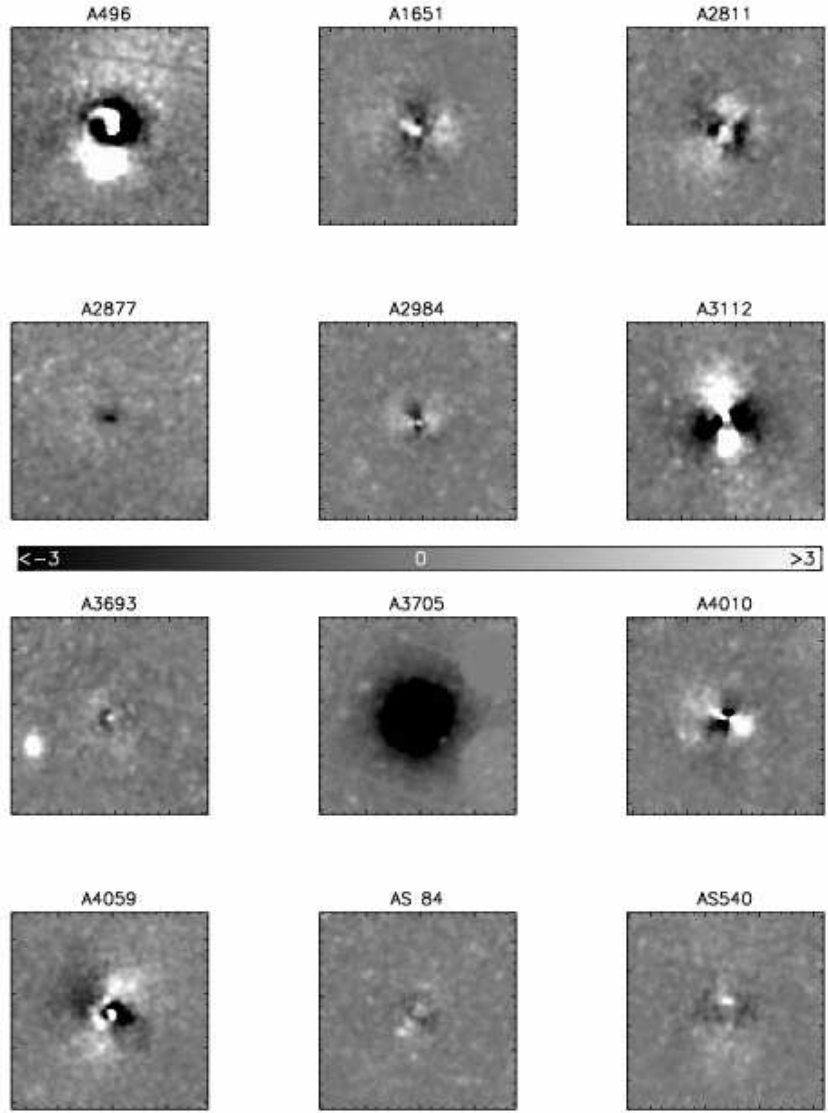}
\figurenum{4}
\caption{$0.3-8.0$ keV 2D surface brightness fit residuals (15\arcmin \ on each side) for observed clusters.  The data are smoothed by a 25\arcsec \ gaussian kernel to make the residuals easier to see. The color bar is in units of counts. These residuals are not normalized by the total observation time. Several cluster residuals have a distinct quadrupole pattern that is an artifact of carrying out a circularly symmetric model fit to an intrinsically elliptical surface brightness distribution. However, several clusters reveal real minor substructure that is clearly asymmetric.  An extreme case is Abell 496, where there is an apparent spiral structure leading to the center of the cluster. Abell 3112 shows an extended tail to the southwest.  Abell 3705's residuals show a very poor fit due to the cluster's high asymmetry and the data's low signal-to-noise. In Abell 3693, there is a bright clump to the southeast. Abell 4059 has a filamentary structure near the center. Overall, the observed substructure does not significantly impact our analyses, as evidenced by the acceptable $\chi^2$ values we obtain for most cluster surface brightness fits.\label{residualfig1} \label{residualfig2}}
\end{figure}

\begin{figure}
\centering
\epsscale{1.1}
\plottwo{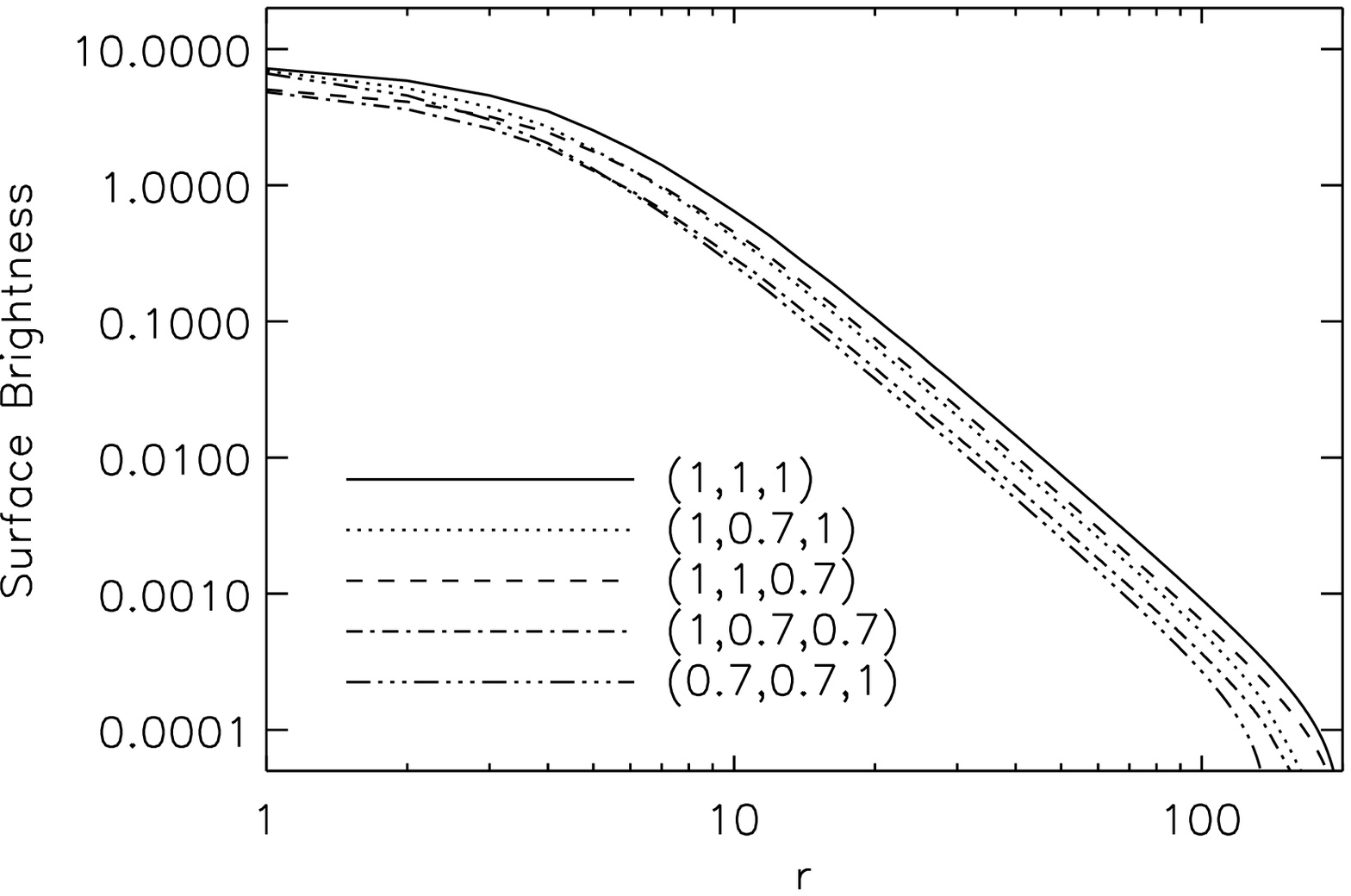}{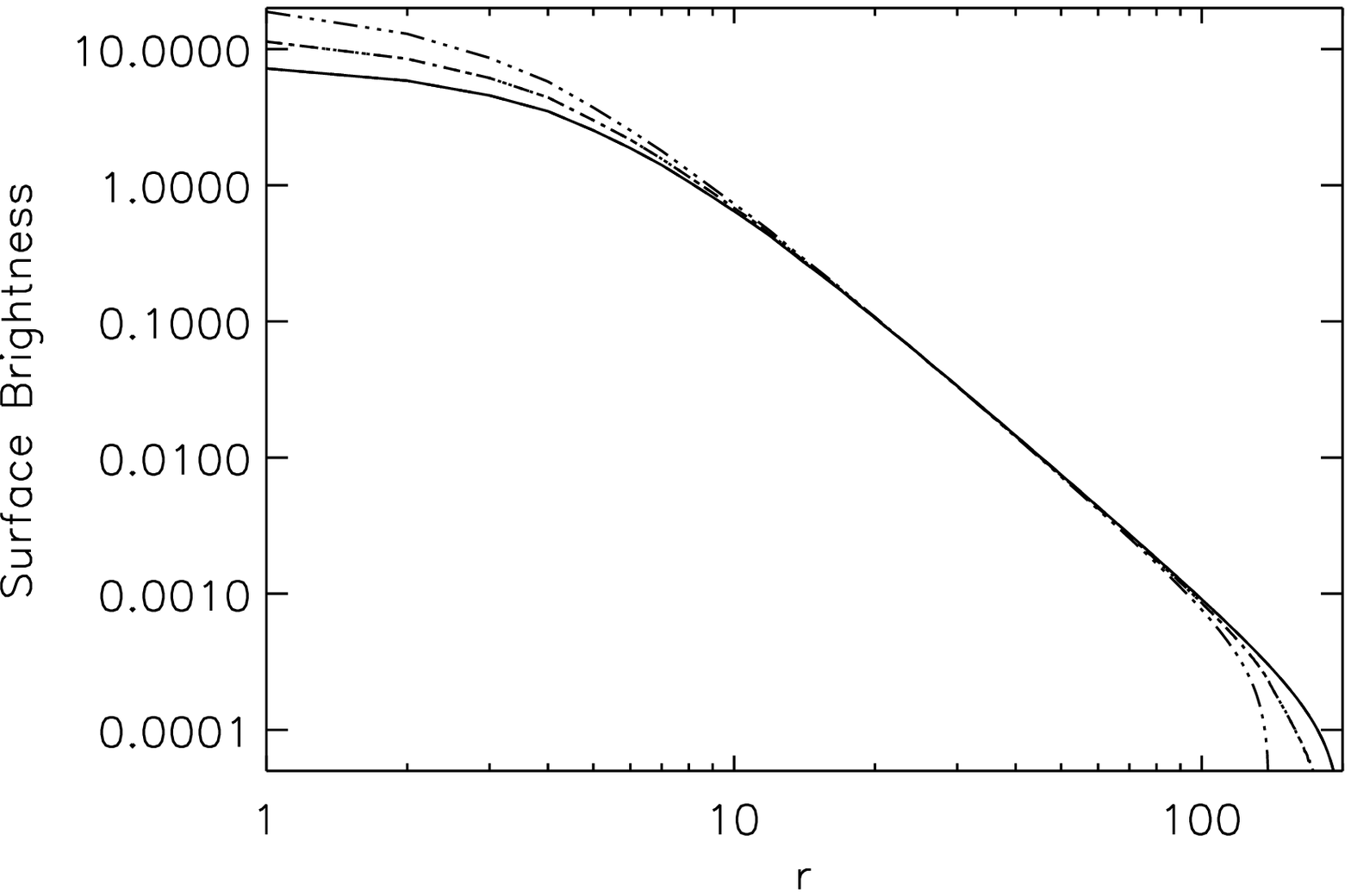}
\figurenum{5}
\caption{
Testing systematic errors arising from the spherical symmetry assumption. In the left
panel we show the azimuthally  averaged surface brightness profile for a range of shapes (the
legend provides $\epsilon_x,\epsilon_y,\epsilon_z$ for each model). Although the different
clusters are offset in surface brightness the profiles are similar in shape. The right panel
demonstrates this to be the case after we normalize the profiles at $r = 25$. These two 
plots suggest that intrinsic shape differences should not lead to dramatic differences in 
profile parameters other than the normalization, except at very large or small radii. The
bias is for elliptical clusters to be fainter than the corresponding spherical cluster.
\label{sphereerror}}
\end{figure}

\begin{figure}
\epsscale{1.1}
\plottwo{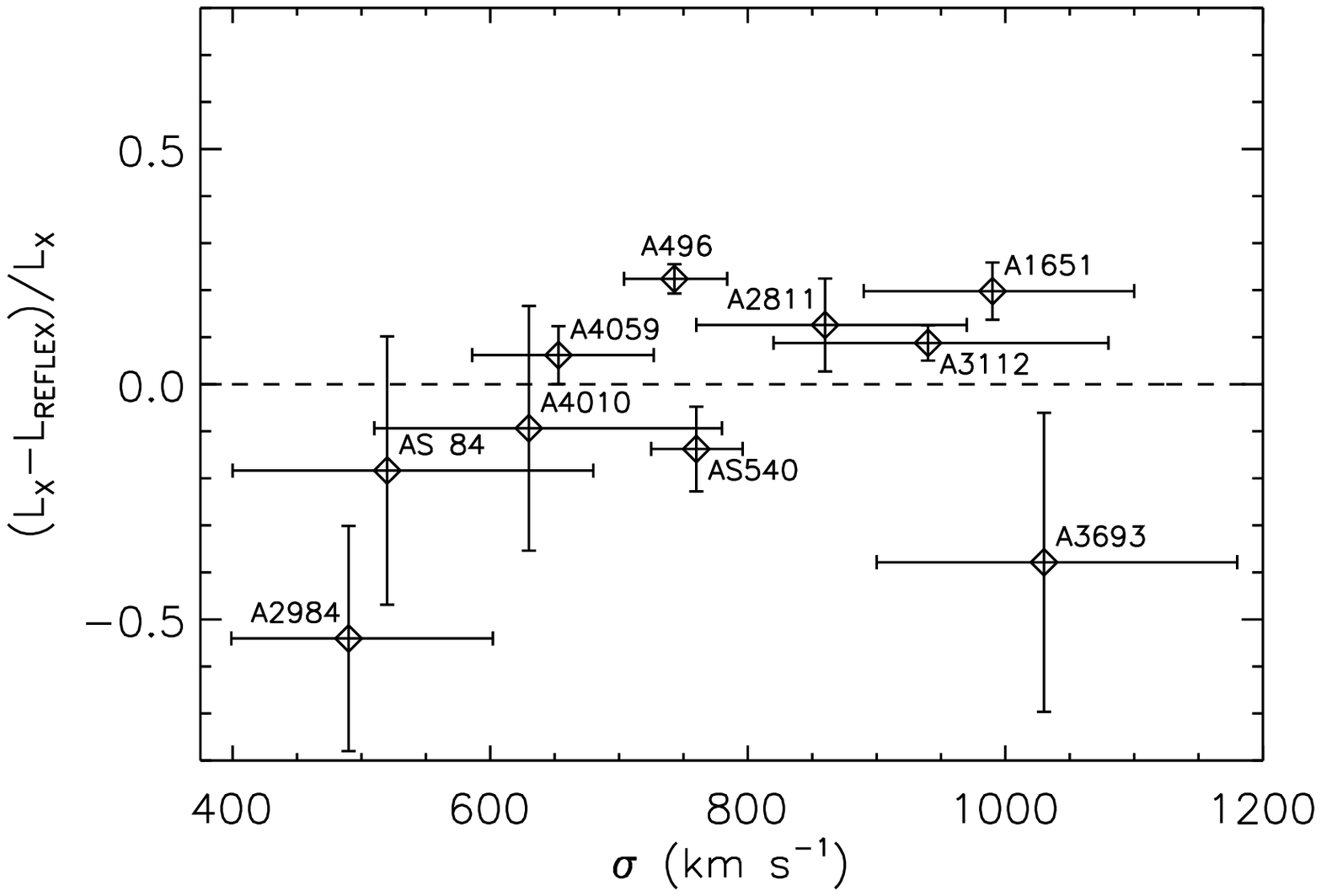}{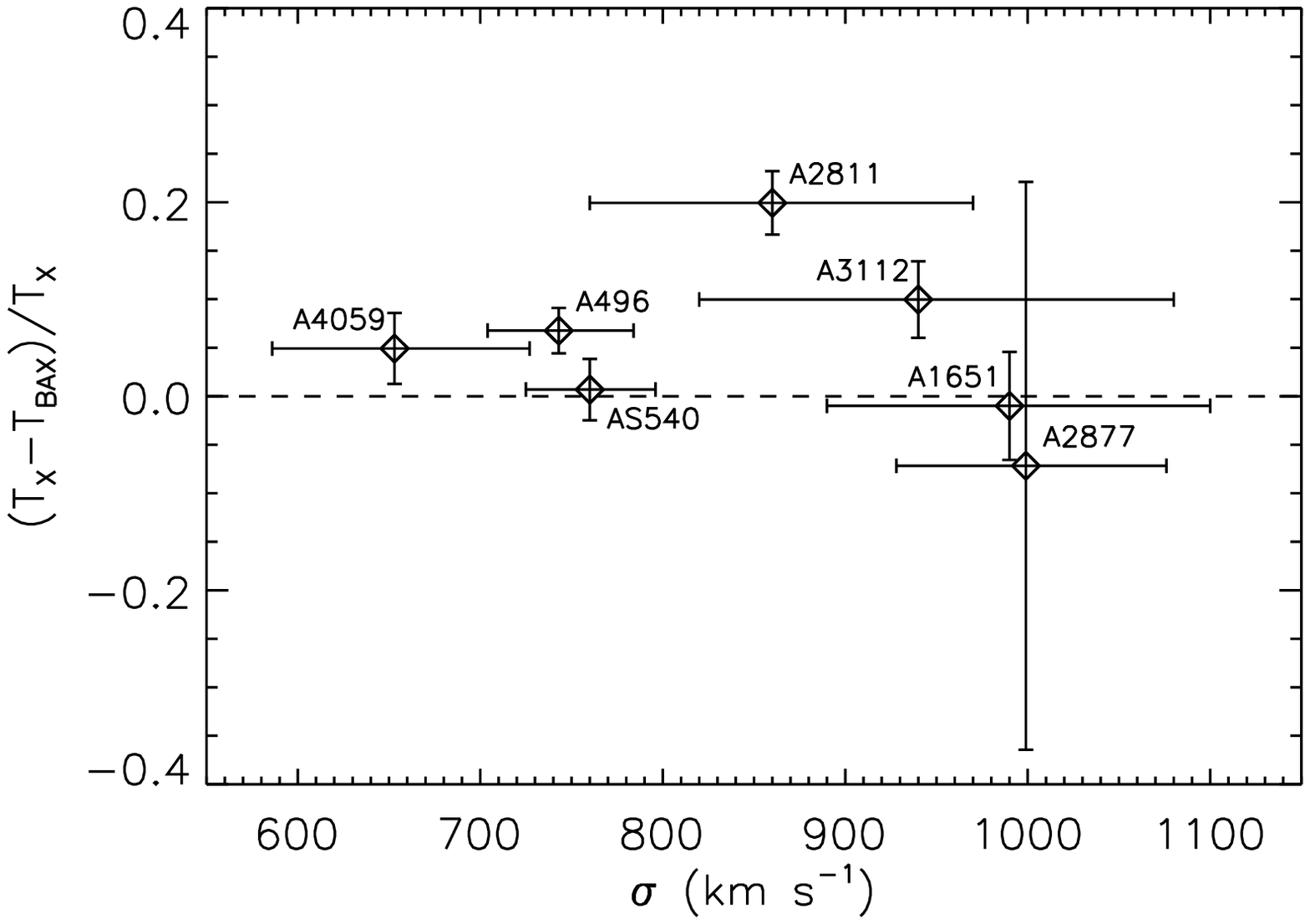}
\figurenum{6}
\caption{
{\it Left:} 
Differences in the $0.1-2.4$ keV luminosities determined by this work ($L_X$) and those from REFLEX \citep[$L_{REFLEX}$;][]{bohringer04} within an aperture defined by the REFLEX ROSAT observations. The 
measurement differences for most clusters fall within the 20\% envelope, although our measurements tend to be brighter than theirs at higher velocity dispersions. 
{\it Right:} Differences in temperatures determined by this work ($T_X$) and those from the BAX database ($T_{BAX}$) for seven clusters. Three out of four clusters (Abell 496, 3112, and 4059) with systematically lower temperatures in BAX have cool-cores. 
We expect the BAX temperatures, typically derived using emission-weighted methods, to be systematically lower than our
peak temperatures in such cases. However, according to our spectral fits, the fourth cluster, Abell 2811, is isothermal, suggesting its discrepancy is real.
\label{rcomp}}
\end{figure}

\begin{figure}
\epsscale{0.75}
\plotone{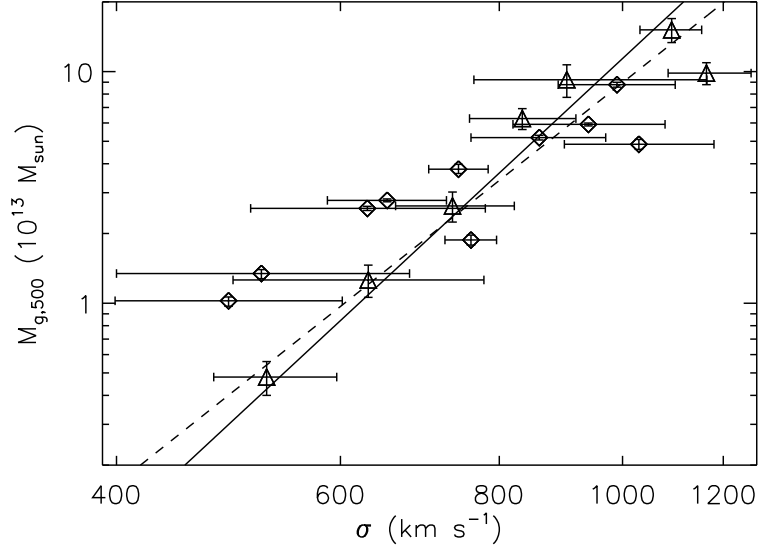}
\figurenum{7}
\caption{Gas mass within r$_{500}$ plotted against cluster velocity dispersion. The diamond points are our clusters and the triangular points are the seven \cite{vikhlinin06} X-ray clusters with known velocity dispersions. The dashed line is a power law fit to the \cite{vikhlinin06}  data, while the solid black line is the fit to our data. The fits agree closely. 
We caution when interpreting results from the lowest velocity dispersion clusters, because the
relationship used to estimate the r$_{500}$ values is poorly constrained in that velocity dispersion
range.
\label{gascomparisons}}
\end{figure}

\begin{figure}
\centering
\epsscale{1.0}
\plotone{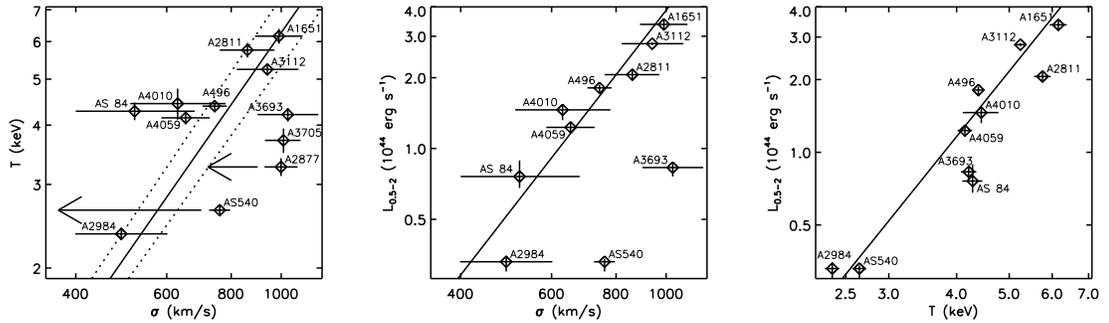}
\figurenum{8}
\caption{{\it Left:} Cluster temperature (derived from peak cluster temperature) versus cluster velocity dispersion. The text beside each datapoint is the cluster's name. The solid line is the T-$\sigma$ fit taken from \cite{wu}; the dotted lines represent 1$\sigma$ errors of the fit. The majority of the clusters lie within the 1$\sigma$ errors of the fit.
The arrows point to alternate velocity dispersions for Abell 2877 (after removing its lower velocity dispersion peak) and Abell S540 (after clipping its velocity wings). 
{\it Middle:} $0.5-2.0$ keV X-ray luminosity plotted against velocity dispersion. The solid line is the best fit excluding the two outliers, Abell 3693 and S540.  
{\it Right:} $0.5-2.0$ keV X-ray luminosity plotted against velocity dispersion. The solid line is the best fit for all clusters. These X-ray-optical relations further establish the validity of the X-ray reduction and analysis techniques employed here.
\label{xraydiag}}
\end{figure}

\clearpage
\thispagestyle{empty}
\begin{figure}
\centering
\epsscale{0.8}
\plotone{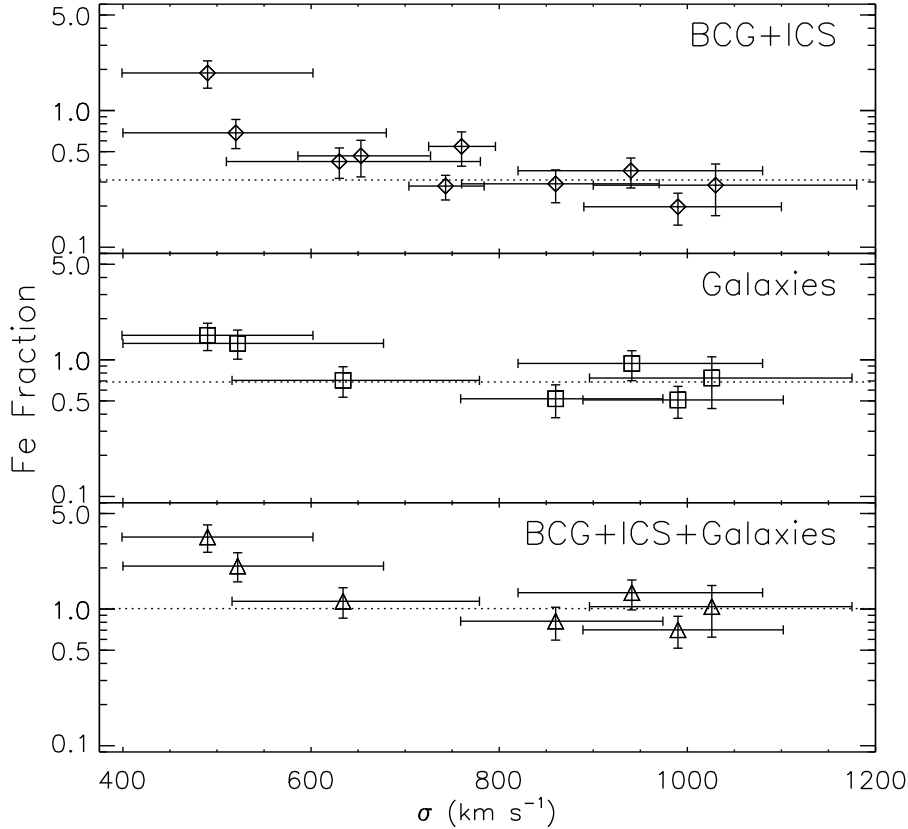}
\figurenum{9}
\caption{{\it Top:} Fraction of the ICM's Fe within $r_{500}$ contributed by
the BCG+ICS component as a function of velocity dispersion. The dotted line at 31\% is the weighted average of the fractional Fe contribution of the full sample. Averages are computed using $1/\textrm{error}^2$ as weights. The Fe fractional contribution changes significantly with velocity dispersion (see text).  However, the r$_{500}$ values for two lowest velocity dispersion clusters (Abell 2984 and Abell S 84) are only constrained by a single low velocity dispersion point in the calibration of the r$_{500}-\sigma$ relation \citep{gonzalez07}, and we suggest caution in interpreting results from these two data points.
{\it Middle:} Fraction of the ICM's Fe within $r_{500}$ contributed by non-BCG cluster galaxies
as a function of velocity dispersion. We assume 84\% metal loss, the value required to generate on average all the ICM's Fe (see bottom panel) from the combined contribution of the BCG+ICS and galactic stellar
components.  We plot only the subset of clusters for which the galactic stellar component is well measured \citep{gonzalez07}. The dotted line at 69\% is the weighted average of the galactic contribution. The galaxies alone cannot account for all the metals in the ICM even if they have a 100\% metal mass loss efficiency. 
{\it Bottom:} Fraction of the ICM's Fe within $r_{500}$ contributed by all the stars
--- in and out of galaxies --- as a function of velocity dispersion. Only clusters with both BCG+ICS and galactic light measurements are plotted. A galaxy metal loss fraction of 84\% is chosen so that the weighted average of the total stellar mass contribution (dotted line) is 1, thereby accounting for all the measured Fe.  While this metal loss fraction is large, it is consistent with some estimates
\citep[$\sim $80\%;][]{calura06} and with the fraction of metals not still locked up
in galactic stars ($\sim$75\%).  If we increase the rate of SN Ia in cluster galaxies by a factor of $1.8$,
to its upper $2\sigma$ bound \citep{mannucci07}, 
we can produce all the ICM's Fe with a galactic metal loss fraction of only $\sim$35\%,
which is consistent with other metal loss fraction estimates \citep{bouche07}.
 \label{finalresult}}
\end{figure}


\clearpage
\appendix
\section{Notes on Individual Clusters}
\subsection{Abell 496}
This is a nearby high signal-to-noise, mildly elliptical, cool-core cluster with emission that overfills the XMM-Newton field-of-view. This complicates surface brightness analysis because the flat cosmic background is masked by cluster emission that is at least an order of magnitude stronger within the field-of-view. The initial surface brightness fits converge at zero background due to the combination of asymmetric structure near the core and bright cluster emission compared to the background. In Figure \ref{residualfig1} we show the spiral-like structure that remains in the central parts of the cluster after subtracting off the best fit beta model. This feature is also present in Chandra exposures of the cluster core. It is suggestive of cooler gas spiraling into the cluster's center, but this can also be explained by chance alignment of higher density gas clumps. We correct the poorly fit background by carrying out a new fit with the central 4\arcmin \ excised. This removes the central asymmetries, resulting in a fit that converges on a non-zero background. We then fix the background in the original surface brightness fit to the newly derived background values and fit again for the beta model parameters. We use these values in our further analyses. We also excise the central 20\arcsec \ of this cluster to remove the brightest cluster galaxy and fit only the extended cluster component of the beta model. Despite our attempts at improving the fit, the surface brightness residual, see Figure \ref{sbspecplot1}, is large, showing a systematic trend with radius. A possible explanation for this is the fainter asymmetric structure observed in the residual north of the cluster center.

\subsection{Abell 1651}
One of the higher redshift and most massive clusters in our sample, this is the best-behaved cluster with reasonable signal-to-noise, even though half of the exposure time was ruined by flaring events. The X-ray contours are circular and are centered on the brightest cluster galaxy. The cluster temperature profile is isothermal and, as expected, its metallicity profile shows no gradient. The surface brightness fit is consistent with the data and shows no systematic variations with radius. Two-dimensional fit residual in Figure \ref{residualfig1} shows a mild asymmetric structure centered around the cluster core, which is an artifact of the circularly symmetric models as the cluster is slightly elliptical. The relative contributions of the BCG and ICS to the BCG+ICS light is poorly understood for this cluster. This is the only cluster in our sample that can be well fit by a single de Vaucouleurs profile \citep{gonzalez00}, making the distinctions between BCG and ICS somewhat arbitrary in the two component de Vaucouleurs fit \citep{gonzalez05}.

\subsection{Abell 2811}
This is another well-behaved cluster whose surface brightness profile is close to circularly symmetric. However, the diffuse cluster emission is offset from the brightest cluster galaxy by 27$^{\prime\prime}$ in the northwest direction. There are no systematic trends observed in the surface brightness residuals. Just like Abell 1651, this cluster's temperature profile is isothermal and exhibits no metallicity gradients. This cluster is a significant outlier in the $L_{X}-T$ relation and appears to be underluminous for its temperature. When we compare to the BAX results \citep{sadat04}, Figure \ref{rcomp}, we see that our derived cluster temperature is higher than the BAX temperature. We also notice that this cluster is hotter than what is expected for its luminosity in our $L_X-T$ relation, suggesting that these discrepancies are real.

\subsection{Abell 2877}
The lowest redshift cluster in this sample, Abell 2877 is one of the least well-behaved clusters. Its surface brightness profile has an uncharacteristically low $\beta$ value of 0.27, and it is not well-fit by a beta profile. The X-ray image shows a very luminous central source surrounded by much fainter extended diffuse emission.  The central source, which is coincident with the brightest cluster galaxy, is offset south from the center of the diffuse emission. We excise a 47$^{\prime\prime}$ radius region around the central source to remove all of its light and fit just the extended emission.  Despite our attempts to remove any sources of problems for our fit, the fit consistently converges to unphysically small core radius and small beta value. A NASA Extragalactic Database\footnotemark[5] (NED) query for this cluster shows that there are two superimposed clusters separated by $\Delta z \sim 0.006$ in redshift space, which is a likely explanation for the poor surface brightness profile fits. A velocity histogram of the member galaxies unsurprisingly shows a second, albeit smaller, peak, likely leading to an overprediction of the cluster velocity dispersion. This cluster is an outlier in the $T - \sigma$ relation, suggesting that
$\sigma$ is indeed overestimated.  Removing the smaller velocity peak from the distribution 
and recalculating $\sigma$ moves
A2877 to within the errors of the relation.
We also center the spectral extraction radii used for our spectroscopic fits about the the extended emission. This cluster exhibits a metallicity gradient, and the abundance falls precipitously at large radii, which is not observed in the other clusters. For these reasons, we remove this anomalous,
apparently unrelaxed cluster from our analysis.

\footnotetext[5]{http://nedwww.ipac.caltech.edu/}

\subsection{Abell 2984}
This is the lowest velocity dispersion cluster in the sample and also the smallest in physical extent (see Figure \ref{contourfig1}). The cluster is not axis-symmetric, though the surface brightness fit does a good overall job fitting the cluster, as indicated by the fit residuals. This cluster does not look peculiar in the various $T-L_X-\sigma$ relations. Its gas mass also appears consistent with that expected from the relation derived from gas mass data from \citet{vikhlinin06}, Figure \ref{gascomparisons}. However, the BCG+ICS makes an unphysical fractional Fe contribution (greater than unity) to the ICM metal budget (see $\S$\ref{metalssec}). 
The origin of this overprediction is unclear.
Figure \ref{rcomp} suggests that we underpredict the luminosity of this cluster, which will in turn lead to an underprediction of gas mass and metal mass. 
It may also be that the intracluster light measurements are flawed and conspire to produce too large a contribution to the ICM metals, or it could be that our scaling of the PEGASE model to match the cluster early-type SN Ia rate may not be applicable for this large group. If we neglect this rescaling, then the BCG+ICS contribution falls to a physically believable value (less than unity). Note that this cluster has both a metallicity gradient and an isothermal temperature profile.

\subsection{Abell 3112}
This cluster is relatively well-behaved, though it is one of the more elliptical clusters in the sample. It has some extended emission from what appears an infalling clump south of the cluster center. This clump, however, has negligible impact on the overall cluster surface brightness distribution and is not evident in the X-ray contour plots. Overall, the beta profile does a good job of describing the cluster surface brightness profile. There is some discrepancy between the fit and the data at small radii, but that is expected when fitting a circularly symmetric model to an elliptical distribution. We do excise the inner 35$^{\prime\prime}$ to remove the cool-core region from the fit. This cluster does not appear anomalous in all of our diagnostics. 

\subsection{Abell 3693}
This is the highest redshift cluster in our sample and has moderate signal-to-noise data. Even though the diffuse cluster emission appears circular, we observe a large clump to the southeast of cluster. This structure is evident in the surface brightness profile and fit residuals at a radius of 900 kpc from the cluster center. The velocity histogram has multiple peaks, suggesting that the velocity dispersion of this cluster may be overestimated. This cluster is a significant outlier in both the $T-\sigma$ and $L_X-\sigma$ relations, which is no surprise, but it is also somewhat discrepant in the $L_X-T$ plot.  One possible explanation is that this system is undergoing a minor merger, which is not likely to significantly impact our analyses.

\subsection{Abell 3705}
This is the least spherically symmetric and lowest signal-to-noise cluster in our sample, making it impossible for a surface brightness fit to converge. In the X-ray contour plot, two clumps of equivalent size dominate the X-ray emission. For a simple circularly symmetric beta profile, the fit parameters converge to senseless values and produce very large fit residuals. Multiple approaches, including excising a large region around the smaller clump to the northwest of the cluster center, fail. For this reason, we remove the cluster from further analysis. However, it is instructive to look at the various diagnostic plots, because we are able to extract a few X-ray parameters from the data. This cluster is an outlier in the $T-\sigma$ plot. Its velocity histogram has multiple peaks hinting at substructure and
suggesting that our $\sigma$ is an overestimate. This unrelaxed system is the sole Bautz-Morgan Type III cluster in our sample. 

\subsection{Abell 4010}
The X-ray data for this cluster is of moderate signal-to-noise.  The surface brightness map looks regular with the exception of a slight asymmetry to the southeast in the last X-ray contour in Figure \ref{contourfig2} and in the surface brightness fit residual 300 kpc from the cluster center. We excise the inner 35$^{\prime\prime}$ to remove the cool-core and fit the extended X-ray component. The beta model fit does a reasonable job in fitting the surface brightness profile. The cluster appears ``normal" in the $T-\sigma$, $L_X-\sigma$, and $L_X-T$ diagnostics.

\subsection{Abell 4059}
We obtained high signal-to-noise X-ray observations of this cluster. The cluster surface brightness distribution appears smooth and is mildly elliptical. The beta model fit shows no systematic trends in the residuals, and the cluster falls within all of the trends found in our diagnostic plots. There is nothing to suggest this cluster is peculiar. We do not excise the central region of the cluster because our beta profile did not show any large residuals in the cluster center.

\subsection{Abell S 84}
Our moderate signal-to-noise data show this cluster to be regular and circular. Its surface brightness fit is well-behaved. There is a hint of substructure 200 kpc southeast of the cluster center, which appears as a small bump in the residuals. This cluster is roughly consistent with the $T-\sigma$ relation, but is underluminous in the $L_X-T$ plane. One possible explanation is that the temperature is overestimated for this cluster. Even though the temperature profile is isothermal, the metallicity profile has a steep gradient. 

\subsection{Abell S540}
The relatively high signal-to-noise data show the highly elliptical nature of this poor cluster. We excise the central 35$^{\prime\prime}$ to remove the cool-core before conducting our surface brightness fit. There is no systematic trend in the fit residuals. Despite the high velocity dispersion, the cluster temperature suggests that this system is a rich group or poor cluster. It is a 3$\sigma$ outlier to the $T-\sigma$ and $L_X-\sigma$ fits. 
The cluster, however, falls on the mean $L_X-T$ trend that we observe for our sample.
The velocity histogram for the member galaxies is poorly-sampled, so it is likely that the velocity dispersion is overestimated. 

\end{document}